\documentclass[aps,pra,onecolumn,nofootinbib,superscriptaddress]{revtex4-2}
\usepackage{graphicx}
\usepackage{dcolumn} 
\usepackage{bm}      
\usepackage{mathtools}
\usepackage{color}
\usepackage{float}
\usepackage{bbm}
\usepackage{amssymb,amsmath,amsthm,amsfonts}
\usepackage{natbib}
\usepackage{circuitikz}
\usepackage{tikz}
\usetikzlibrary{patterns}
\usepackage{hyperref}
\usepackage{tikz-cd}
\newtheorem{theorem}{Theorem}[section]

\begin{document}

\title{Topological phases of matter, Quantum Error Correction and Topological twists}

\author{Tushar Pandey}
\email{tusharp@tamu.edu}
\affiliation{Department of Mathematics, Texas A\&M University, College Station, TX 77840}

\begin{abstract}
    Unitary Modular Tensor Categories(UMTC) have a one-to-one correspondence with Topological Quantum Field Theories (TQFT). Different identifications have been made so far associating different physical particle types (anyons) to different UMTCs. However, the area of anyon model transformations has not been discovered much, which corresponds to topological phase transitions. Physical condensation of particles (quasi-particles) can correspond to functors between different UMTCs. This paper discusses a special case of doubled semion condensation. We also give an error detection and correction algorithm for the Doubled Semion model which is more efficient than the previously suggested ones. At last, we present some results about twisting the underlying lattice structure and see its effects on UMTCs and logical subspaces e.g., for Ising, Doubled Semion models, and a combination of these. This suggests some interesting "wormholes" phenomenon.
\end{abstract}

\maketitle

\section{Introduction}
A few decades ago, Topological Quantum Field Theories were discovered and a relationship between them and Unitary Modular Tensor Categories was formulated. Later, an identification was found between TQFTs and UMTCs which provided a strong foundation for studying Topological Phases of Matter \cite{Rowell2007} (TPM) and Topological Quantum Computation (TQC). TQC provides a more robust and fault-tolerant approach towards Quantum Computation with appropriate TQFT in the background. Different TQFTs (and UMTCs) correspond to different possible computations, with the Ising model/category being capable of performing most gate operations (Clifford gates). With an additional T gate, Ising UMTC can do \textit{Universal Quantum Computing} (which can also be done just with Fibonacci anyons/model).
\\
The practical aspects of these topological quantum theories are still limited. The only TQFT + TQC that has been tested is the Toric code \cite{Satzinger2021} (and Surface codes). However, the computational power of such models is not strong enough for computing. They are good for error correction, depending on the perspective. A recent result \cite{Acharya2022} by a team at Google showed an increase in the error correction efficiency by growing the lattice size for surface codes. While the error correction seems promising, there are ways to change the structure of the underlying lattices to create exciting properties, including new particle-like behavior \cite{Bombin2010}. Breaking the structure of the lattice at the boundary also creates interesting results \cite{Kitaev2012}.
\\
There are some ways to get to other models, using a higher dimensional Toric code embedding \cite{Ellison2022}. The categorical approach to the same \cite{Kong2014} creates the possibility of developing tools to look into the phase transitions through mathematics. In this paper, we discuss twists on the $\mathbb{Z}_2$ Toric code, relate that to creating wormholes \cite{Krishna2020} in possibly a pair of Toric codes on two levels, and look into a different picture of \cite{Bombin2010}'s model. We also discuss the statistics and computational logistics of the new models. These twists can also be seen geometrically with some topological implications related to local and global twisting. We formulate the error detection and therefore correction algorithms for the Doubled Semion model explicitly. We give a mathematical formulation to describe \cite{Ellison2022}'s condensation. With some rough ideas, we let the readers ponder over the path toward experimental verification of this phenomenon.
\\
\\
\textbf{Outline of the paper}
\\
In section II, we revise the basics of category theory and Toric codes as preliminaries. We look into different UMTCs and Topological phases of matter. We also discuss $\mathbb{Z}_n $ Toric codes in the same section. In section III, we look into twists and see some topological implications. We review Bombin's twist, generate a dual version of the same, and then generalize it for higher twist structures. In section IV, we discuss QECC for the doubled semion model explicitly. Next, we discuss Topological phase transitions between different models from Toric codes. We review some of the work \cite{Ellison2022}, \cite{Bombin2010}, \cite{Kitaev2012} relating the Toric Code topological order into Doubled Semion and Ising topological order. We look into the possibility of making wormholes of different kinds through the ideas of twists and toric code couplings.
\\
\\
\\
\textbf{Acknowledgements}
\\
I would like to thank Eugene Dumitrescu for important comments, physical interpretations, and ideas. This research was supported in part by an appointment with the NSF MSGRI Program at ORNL. I also acknowledge DOE ASCR funding under the Quantum Computing Application Teams program, FWP number ERKJ347. The research aligned with the goals of the Quantum Science Center at ORNL.

\section{Preliminaries}
\subsection{Categories and Functors}
The notations for this section are similar to \cite{Kong2014} Appendix.
The isomorphisms mentioned in the section are technically natural isomorphisms as in the language of category theory.

A category $\mathcal{C}$ consists of labels as objects and arrows as morphisms. For example, the category of sets, \textbf{Set}, has `sets' as objects and `functions' as morphisms. If there is a morphism from A to B and one from B to C, then there is a morphism from A to C (defined by composition), where A, B, and C are the objects in $\mathcal{C}$. A priori, categories do not have structures, so different structures can be chosen. One such structure is a tensor product or monoidal structure.
\\
\\
\textbf{Monoidal category (Tensor category)} comes with a tensor product $\otimes$ and a tensor unit $\bm{1}$ (corresponds to vacuum in physics). The tensor product of two objects should also be an object in the category. Similarly, there is a notion of tensor product on morphisms (like the usual one) $ f\otimes g: X \otimes Y \to Z \otimes W$ would mean, f is a morphism from X to Z and g is a morphism from Y to W. There are certain axioms that the tensor product maps need to satisfy. There is an associative isomorphism for the tensor product
\begin{align*}
    \alpha_{X,Y,Z}: (X \otimes Y) \otimes Z \xrightarrow{} X \otimes (Y \otimes Z)
\end{align*}
which satisfies the \textit{pentagon axiom} \cite{Delaney2019}. There are unit isomorphisms which we will omit for this paper.
\\
One can also add a structure called braiding (\textit{Braided monoidal category}) which is an isomorphism $c_{X, Y}: X \otimes Y \rightarrow Y \otimes X$ and together with $\alpha$ isomorphisms, they satisfy \textit{hexagon relations} \cite{Delaney2019}.
\\
\\
A \textit{$\mathbb{C}$-linear category} is a category where the hom spaces are vector spaces over $\mathbb{C}$. A category is called \textit{semi-simple} if any object is a direct sum of finitely many simple objects, $A = \oplus_{i=1}^k X_i $ where $X_i $ is a simple object for some $k$.
\\
\\
We want the unit $\bm{1}$ to be a simple object (\textit{Fusion category}).
Oftentimes, one would encounter F-matrices in the intersection of physics and category theory, which corresponds to a change of basis coefficient and is also related to angular momenta(recoupling) coefficients. It's equivalent to the associativity isomorphism between two morphisms.
\begin{equation}\label{change-of-basis}
    hom((X \otimes Y) \otimes Z, W) \xrightarrow{F_{XYZ}^W} hom(X \otimes (Y \otimes Z), W)
\end{equation}
where $F_{XYZ}^W$ is a matrix with dimensions depending on $X \otimes Y$ and $Y \otimes Z$. It's also known as fusion matrices in physics.
\\
A category is called \textit{Rigid} if any object (X) has both left(X*) and right(*X) duals, with birth (coev) and death (ev) maps defined by:
\begin{align} \label{rigidity maps}
    d_X: X^{*} \otimes X \rightarrow \bm 1, \;\;\;\;\;\;\;\;\;\; d_X^*: X \otimes {}^* X \rightarrow \bm 1
    \\
    b_X : \bm 1 \rightarrow X \otimes X^{*}, \;\;\;\;\;\;\;\;\;\; b_X^*: \bm 1 \rightarrow {}^* X \otimes X
\end{align}
For a skeletal category (smallest equivalent subcategory of the original category), one can do graphical calculus with strings and boxes, and realize these maps via string diagrams with morphisms represented by boxes.
\begin{figure}[h]
    \centering
    \includegraphics[width = 0.45\textwidth]{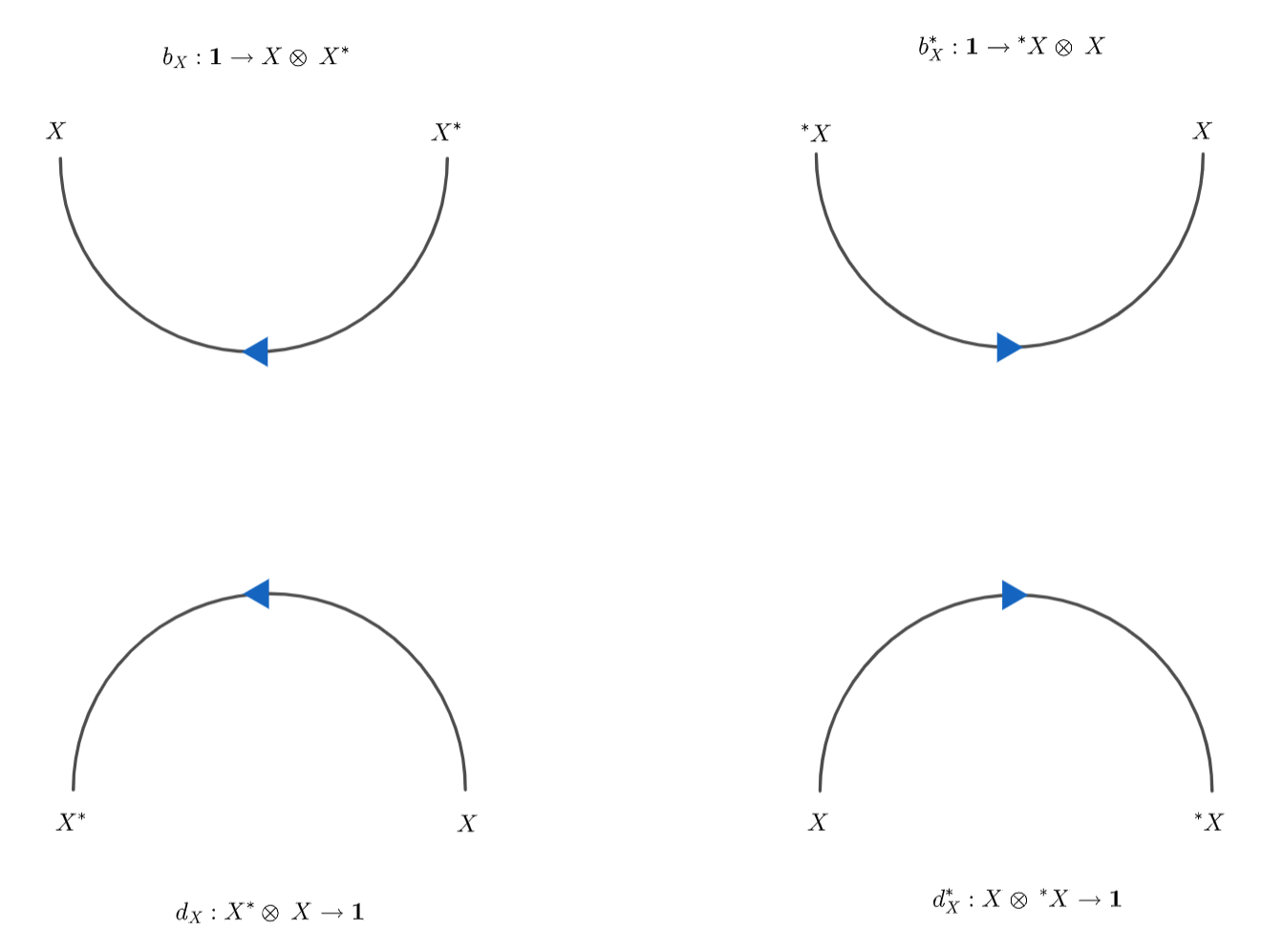}
    \includegraphics[width = 0.5 \textwidth]{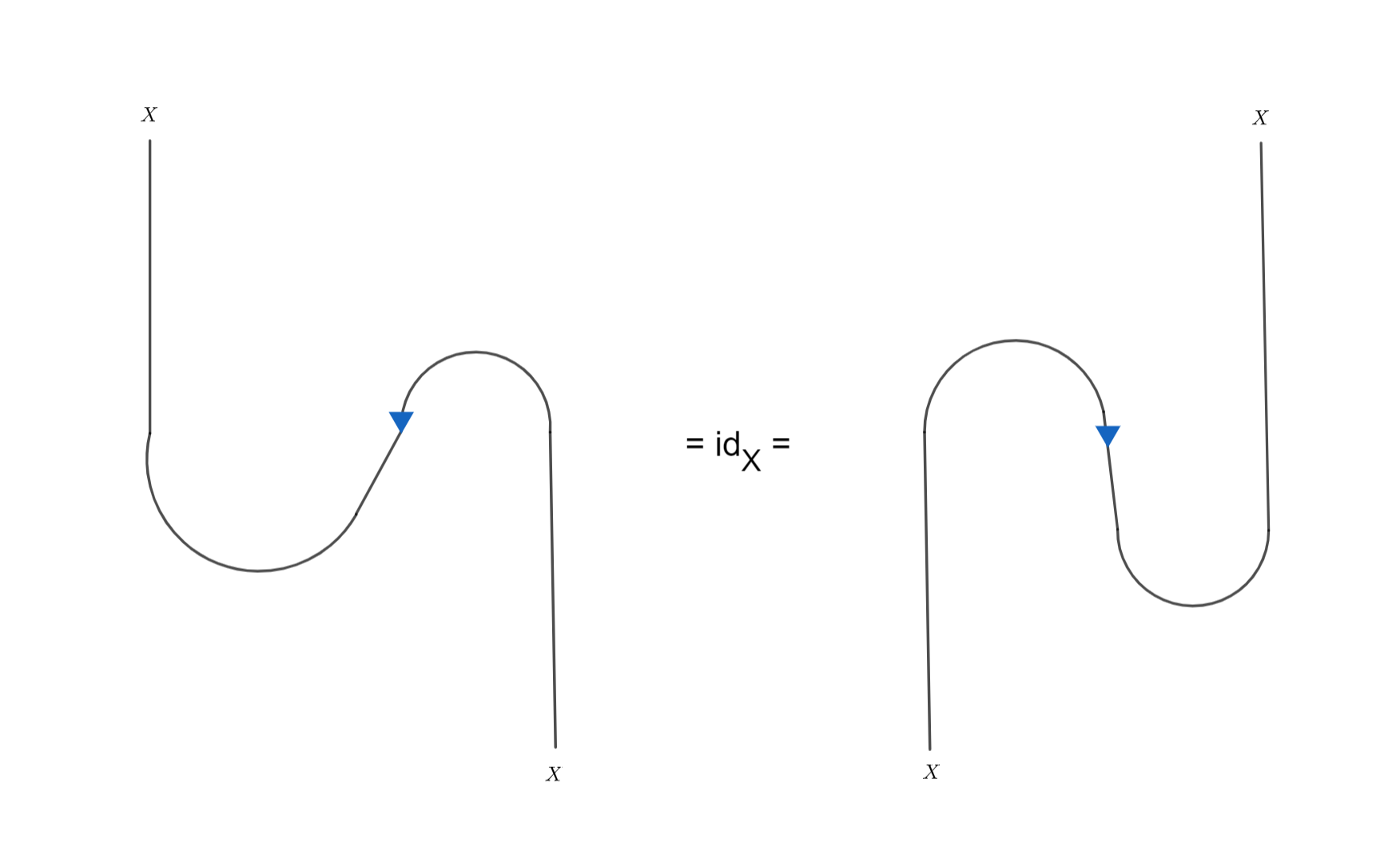}
    \caption{String diagrams for skeletal categories}
\end{figure}
\\
A \textit{Pivotal structure} means the dual of an object's dual, is the object itself (up to isomorphism), i.e. $X \simeq (X^{*})^* \simeq {}^*({}^* X) $. For a morphism `f' between $X$ and $(X^*)^*$, $f \in hom(X, (X^*)^*)$ one can define a left trace (a right trace for $g \in hom(X, {}^*({}^*X))$) as
\begin{align}\label{trace maps}
    Tr^L: \bm 1 \xrightarrow{b_X} X \otimes X^* \xrightarrow{f \otimes \bm 1} (X^*)^* \otimes X^* \xrightarrow{d_{X^*}} \bm 1
    \\
    Tr^R: \bm 1 \xrightarrow{b^*_{{}^*X}} {}^*X \otimes X \xrightarrow{\bm 1 \otimes g } {}^*X \otimes {}^*({}^* X) \xrightarrow{d^*_{{}^*X}} \bm 1
\end{align}
A pivotal category is called \textit{ Spherical} if $Tr^L = Tr^R$. For a spherical category, the quantum dimension of X, $dim (X) = Tr(f)$, where $f \in hom(X, (X^*)^*)$. The quantum dimension of a category then is defined as $\mathcal{D} = \sqrt{\sum_X dim(X)^2} $ where the sum is over all simple objects. For a unitary fusion category, the quantum dimensions of objects are real and positive \cite{Etingof2002}.
\\
\\
A \textit{Ribbon category} is a rigid braided tensor category with a twist $\theta_X: X \xrightarrow{\simeq} X $ isomorphism satisfying:
\begin{equation}\label{twist}
    \theta_{\bm 1} = id_{\bm 1}, \;\;\;\;\;\;\;\; \theta_{X^*} = (\theta_X)^*, \;\;\;\;\;\;\;\;\;\;\; \theta_{X \otimes Y} = c_{Y,X} \circ c_{X,Y} \circ (\theta_X \otimes \theta_Y)
\end{equation}
A \textbf{Unitary Modular Tensor Category} is a $\mathbb{C}$-linear semi-simple unitary ribbon category such that we have a non-degenerate S matrix and a diagonal T matrix, where $s_{a,b}$ and $\theta_a$ are defined as:
\begin{figure}[h]
    \centering
    \includegraphics[width = 0.2\textwidth]{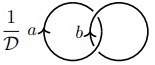} \;\;\;\;\;\;\;\;\;\;\;\; \includegraphics[width = 0.2\textwidth]{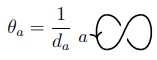}
    \caption{Modular data presentation}
\end{figure}
\\
The first row (therefore first column) of the S matrix describes the quantum dimension of each element because it means taking a closure of the string which is the trace of itself, and thus, equal to the quantum dimension.
\\
\\
\textbf{Example of UMTC}:
\\
$D(\mathbb{Z}_2)$ or Toric Code. The objects are $\{1, e, m, em\}$.
\\
Fusion rules:
\begin{equation} \label{toric_code_fusion}
    e \times m = em,\;\; e \times em = m, \;\; m \times em = e, \;\; e\times e = 1 = m \times m = em \times em
\end{equation}
The individual dimensions of these are: $dim(1) = dim(e) = dim(m) = dim(em) = 1$. Total quantum dimension $\mathcal{D} = 2$
\\
\\
A Functor is a mapping between two categories sending objects to objects and morphisms to morphisms.
\begin{equation*}
    \begin{tikzcd}[ampersand replacement=\&]
    A \arrow[r] \arrow[d, "f"'] \& \mathcal{F} (A) \arrow[d, "\mathcal{F}(f)"] \\
    B \arrow[r]                 \& \mathcal{F} (B)
    \end{tikzcd}
\end{equation*}
where $A,B \in \mathcal{O}b(\mathcal{C}), \;\;\;\; f \in Hom(A,B), \;\;\;\;\;\; \mathcal{F}(A), \mathcal{F}(B) \in \mathcal{O}b(\mathcal{D})$, \;\;\;\; $\mathcal{F}(f) \in Hom(\mathcal{F}(A), \mathcal{F}(B))$.
\\
We need the functor to map the unit to the unit in respective categories, i.e. $\mathcal{F}(id_A) = id_{F(A)} $. A monoidal functor is defined as a pair $(\mathcal{F},\mathcal{J})$ where $\mathcal{J}_{X,Y}: \mathcal{F}(X) \otimes \mathcal{F}(Y) \rightarrow \mathcal{F}(X \otimes Y) $ is an isomorphism and $\mathcal{F}$ is a functor as defined above.

\subsection{Toric Code}
Toric code is a simple spin-lattice model, defined on a periodic lattice exhibiting a topological order and anyonic excitations. It corresponds to $\mathbb{Z}_2 \times \mathbb{Z}_2$ topological order. It was initially studied and developed by Kitaev \cite{Bravyi98}. It has been described concisely in \cite{Herringer20}. It's one of the first models to be verified by \cite{Satzinger2021} that exhibits the modular data as well as can be used for some computation (for a very limited amount).

Start with a 2D lattice, with periodic boundary conditions, i.e. we identify the left and the right boundary as well as the top and the bottom boundary so that the lattice forms a Torus. Each edge has one qubit (which is a spin 1/2 particle). We start with a
Hamiltonian:
\begin{equation}\label{Toric_code_hamiltonian}
    H = -\sum_v A_v -\sum_p B_p
\end{equation}
where $A_v$ and $B_p$ are operators defined on vertex and plaquettes (faces) respectively.
\begin{figure}[h]
    \centering
    \includegraphics[width = 0.315 \textwidth]{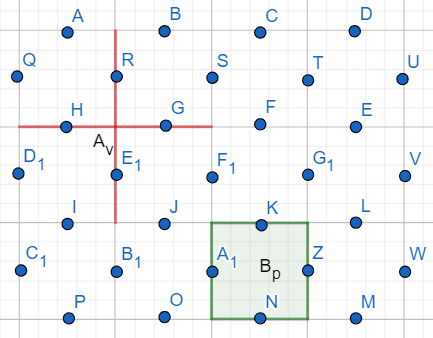}
    \caption{Toric code}
    \label{fig: Toric code}
\end{figure}
\\
More precisely in fig \ref{fig: Toric code}, the red $A_v = \sigma^x_H \sigma_G^x \sigma_R^x \sigma_{E_1}^x$ and the green $B_p = \sigma^z_Z \sigma^z_N \sigma^z_{A_1} \sigma^z_K$. Here $\sigma^x$ and $\sigma^z$ are the Pauli X and Pauli Z operators. In general,
\begin{equation}\label{TC Operators}
    A_v = \bigotimes_{e \in n(v)} \sigma_e^x \;\;\;\;\;\;\;\;\; B_p = \bigotimes_{e \in \partial(face\; p) } \sigma^z_e
\end{equation}
where n(v) is the qubits connected to v (neighbors) and $\partial$(face p) corresponds to the boundary of the face p. They are tensor products of the individual Pauli operators. Both $A_v$ and $B_p$ are commuting Hermitian with eigenvalues, $\pm 1$. The ground state is the +1 eigenstate of each $A_v$ and $B_p$ operator.

\begin{figure}[h]
    \centering
    \includegraphics[width = 0.35 \textwidth]{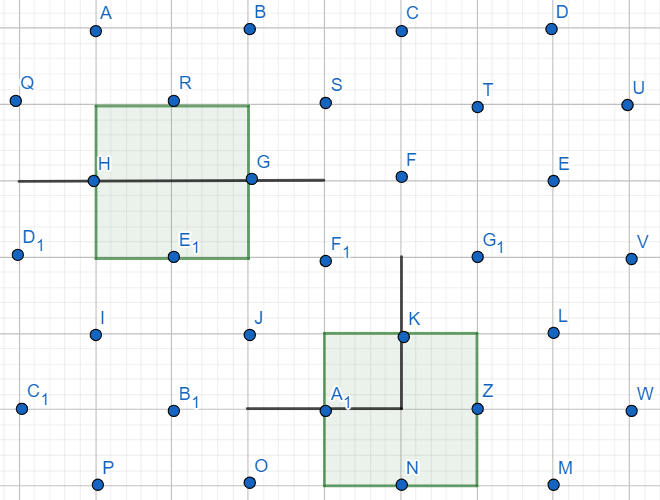} \;\;
    \includegraphics[width = 0.35 \textwidth]{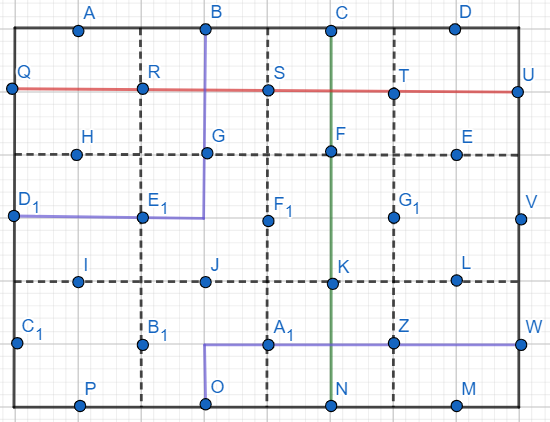}
    \caption{(i) Ground state configurations and (ii) logical operators}
    \label{fig: ground_state_prep}
\end{figure}

\begin{figure} [h]
    \centering
    \includegraphics[width = 0.35 \textwidth]{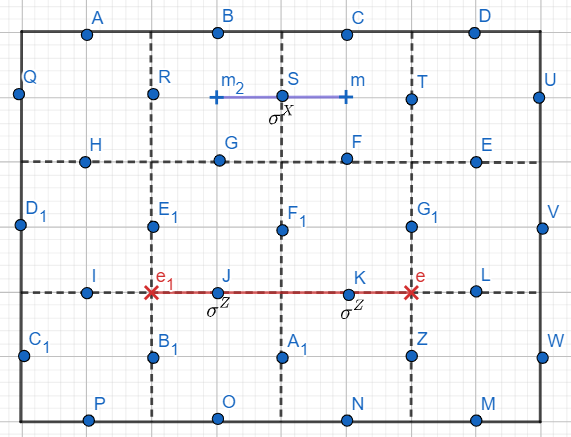}
    \caption{Excitations in Toric code}
    \label{fig: Toric code excitations}
\end{figure}

Consider each unit cell (in the lattice) with two spins on the top left, and consider translations in the horizontal and vertical directions. If there are N translations in the horizontal direction and M translations in the vertical direction, there are 2MN number of qubits. Each qubit is a spin 1/2 particle, therefore the dimension of the Hilbert space is $2^{2MN}$. There are MN number of plaquettes and vertices, therefore 2MN constraints of order two. However,     $\prod_p B_p = 1 = \prod_s A_v$
Therefore the dimension of the ground state is
\begin{equation}\label{dimension TC}
    2^{2MN}/2^{2MN - 2} = 2^2 = 4
\end{equation}
This is also known as the code space and represents the dimension of the logical qubits.
\\
\\
Here, $|0\rangle$ is an eigenvector with eigenvalue +1 and $|1\rangle$ is an eigenvector with eigenvalue -1 for the Pauli Z gate.
\begin{equation}\label{computational basis}
    Z|0\rangle = |0\rangle, \;\;\;\; Z|1\rangle = -|1\rangle; \;\;\;\;\;\;\;\;\;\;\;\; X|0\rangle = |1\rangle, \;\;\;\; X|1\rangle = |0\rangle
\end{equation}
If the state of the qubit is $|1\rangle$, we say it's occupied, and unoccupied otherwise. In order to find the +1 eigenstate for each operator, look at the $B_p$ operator. Four edges have an overall +1 eigenvalue, i.e. even number of them have to be occupied. Consider strings in the dual lattice, based on whether the qubit is occupied or not. If it is, then imagine a string orthogonal to the edge, joining the center of the plaquettes it's shared with. See fig \ref{fig: ground_state_prep}
\\
All the virtual strings in the dual lattice must close out and therefore the ground state has loops. $A_v$ operators change the "occupancy" of the strings. In a vacuum, they create four strings that form a loop. If there was one string already, it extended the loop (made it bigger) and if three strings were occupied, contracted the loop (made it smaller). If two strings were occupied, it either merges two loops or changes the loop (homotopic). Therefore, the ground state, which needs to be a +1 eigenstate of all the operators, is a superposition of all loop configurations on the dual lattice \cite{Bravyi98, Lin2021}.
\begin{equation*}
    |\phi \rangle_{GS} = \bigoplus_{loops} |\text{loop config} \rangle
\end{equation*}
There are some loops that are in the +1 eigenstate but can not be generated by the operators. These are the non-contractible loops on Torus, one along the meridian, one along the longitude, and one like a dehn twisted loop.
There are four degenerate classes of loops on a Torus, explaining the fourfold degeneracy of the code space.
\\
\underline{Excitations}:
\\
The excitations of different models correspond to different UMTCs in the category theory or topological phases of matter. Toric code has three excitations, `e', `m', `$\epsilon$', where $\epsilon = em$. The fusion rules are listed in the example \ref{toric_code_fusion}. There is a benign way to create these excitations through string operators. If one qubit is acted upon by a Pauli X operator, then the state is flipped, which creates a pair of plaquette excitations. Changing the state to a -1 eigenstate causes a change in energy, which is called an excitation. Acting on a qubit with a Pauli Z operator creates a pair of vertex excitations. The string operators are defined by:
\begin{equation} \label{toric_string_op}
    W_P^e = \prod_{e \in P} \sigma^z_e \;\;\;\;\;\;\;\;\;\;\;\; W_{P'}^m = \prod_{e' \in P'} \sigma_{e'}^x
\end{equation}
where P is a path on the lattice and P' is a path on the dual lattice. These are called string operators because they create excitations at the end (of the string/path). See fig \ref{fig: Toric code excitations}
\\
Mathematically, the anti-commutative property of Pauli Z and X causes the excitation.
\begin{equation}\label{toric_excitation}
    B_p \sigma^x |\psi \rangle = -\sigma^x B_p |\psi \rangle = -\sigma^x |\psi \rangle \;\;\;\;\;\;\;\;\;\;\;\;\; A_v \sigma^z |\psi \rangle = -\sigma^z A_v |\psi \rangle = -\sigma^z |\psi \rangle
\end{equation}
where $|\psi \rangle$ was the ground state, i.e. +1 eigenstate of each operator. These errors/excitations can travel for free. Assume there is a path P, with excitations at the end. If we add another Pauli X (or a string of Pauli X), then we get a new path P', and the excitation is only at the new endpoints now, i.e. the energy of the system is the same and only the location of the excitation has changed.
\\
This makes it easier (in some sense) to do error correction. One only needs to know the excitation location to correct them.
This is true until the error propagates across the non-contractible handle, where it changes the logical information. The error cancels itself out but at the cost of a non-trivial loop, therefore changing the loop class of the current system.
\\
\\
\textit{Toric Code and UMTC}: Toric code corresponds to $Z(Rep(\mathbb{Z}_2))$ which is the Drinfield center of the fusion category $Rep(\mathbb{Z}_2)$. The S and T matrices are given by:
\begin{equation}\label{TC modular data}
S :=
    \begin{pmatrix}
    1 & 1 & 1 & 1
    \\
    1 & 1 & -1 & -1
    \\
    1 & -1 & 1 & -1
    \\
    1 & -1 & -1 & 1
    \end{pmatrix}
    \;\;\;\;\;\;\;\;\;
T: =
\begin{pmatrix}
    1 & 0 & 0 & 0
    \\
    0 & 1 & 0 & 0
    \\
    0 & 0 & 1 & 0
    \\
    0 & 0 & 0 & -1
    \end{pmatrix}
\end{equation}
\\
\textit{Fault Tolerant}: For an error to change the quantum information, it needs to propagate all the way around the non-contractible handles, which makes it a topological error. For any local error and error propagation, Toric code is fault tolerant. The error correction algorithm is known for the Toric code, making it a quantum error correcting code. The encoding of information can be done in this way:

\subsection{Topological Phases of matter and Toric Codes}
"What are topological phases of matter? First, they are phases of matter at zero temperature.
Second, they have a non-zero energy gap for the excitations above the ground state. Third, they
are disordered liquids that seem to have no features. But those disordered liquids actually can have
rich patterns of many-body entanglement representing new kinds of order." \cite{Wen2017}
\subsubsection{UMTCs and TQFTs}
Topological phases of matter are phases of matter whose low energy physics is modeled by Topological Quantum Field Theories (TQFT) and their enrichment \cite{Rowell2017}. For more details on topological phases of matter definitions, look into \cite{Rowell2017}, \cite{Wen2017}. This section will describe some relationships between 2D Topological Phases of Matter and Unitary Modular Tensor Categories. Let's look at a table (dictionary) for translation of jargon between physics and math \cite{Rowell2017}.
\begin{table}[h]
    \centering
    \begin{tabular}{|c|c|}
     \hline
     \textit{UMTC} & \textit{Anyonic systems}
     \\
     \hline
     simple objects & anyon
     \\
     \hline
     label & anyon type/topological charge
     \\
     \hline
     tensor product & fusion
     \\
     \hline
     fusion rules & fusion rules
     \\
     \hline
     coefficient in fusion rules $N_{ab}^c$ & fusion space
     \\
     \hline
     dual & antiparticle
     \\
     \hline
     birth/death map & creation/annihilation
     \\
     \hline
     F matrices & recoupling rules
     \\
     \hline
     twist $\theta$ & topological spin
     \\
     \hline
     morphism & physical process/operator
     \\
     \hline
     quantum invariants & topological amplitudes
     \\
     \hline
\end{tabular}
    \caption{UMTC and TQFTs}
    \label{Table 1:}
\end{table}
\\
The elementary excitations gives us the UMTCs and the topology of spaces along with the ground state dependence gives us the Unitary Topological Modular Functors. Some examples of UMTCs and different physics systems corresponding to Topological Phases of Matter.
\\
\\
 $\bullet \;\; D(\mathbb{Z}_2) \text{ or Toric Code }: $
\\
$Simple\;\;\mathcal{O}bj: \{ 1,e,m,em\}$
\\
Fusion: $ e \otimes e = m \otimes m = em \otimes em = 1, \;\; e \otimes m = em \;\; e \otimes em = m \;\; m \otimes em = e$
\\
Quantum Dimension = 1 (for each simple object)
\\
Twists: $\theta(e^p m^q) = (-1)^{pq}$
\\
\\
 $\bullet \;\; D(\mathbb{Z}_4) \;\; or \;\; \mathbb{Z}_4 \text{ Toric code: } $
\\
Simple $\mathcal{O}bj:  \{1,e,e^2, e^3, m, em, e^2m, e^3m , m^2, em^2, e^2m^2, e^3m^2, m^3, em^3, e^2m^3, e^3m^3 \} $
\\
Fusion: similar to Toric code, $e^a m^b \otimes e^c m^d = e^{a + c \; mod \; 4 } m^{b + d \;\; mod\; 4}$
\\
Quantum dimension = 1 (for each simple object)
\\
Twists: $\theta(e^p m^q) = i^{pq}$
\\
\\
 $\bullet \;\; D(\mathbb{Z}_N) \;\; or \;\; \mathbb{Z}_N \text{ Toric code: } $
\\
Simple $\mathcal{O}bj:  \{e^a m^b: 0 \leq a,b \leq N-1\} $
\\
Fusion: $e^a m^b \otimes e^c m^d = e^{a + c \; mod \; N } m^{b + d \;\; mod\; N}$
\\
Quantum dimension = 1 (for each simple object)
\\
Twists: $\theta(e^p m^q) = e^{2\pi i pq / N}$
\\
\\
$\bullet \;\;$ Semion MTC:
\\
Simple $\mathcal{O}bj: \{ 1, s \}$
\\
Fusion: $s \otimes s = 1$
\\
Quantum dimension: 1 (for each simple object)
\\
Twists: $\theta(s) = i$
\\
\\
$\bullet\;\;$ Doubled Semion MTC:
\\
Simple $\mathcal{O}bj: \{1, s, \Bar{s}, s\Bar{s}\}$
\\
Fusion: $s \otimes s = \bar{s} \otimes \bar{s} = s\bar{s} \otimes s\bar{s} = 1$, \;\; $s \otimes \bar{s} = s\bar{s}, \; s \otimes s\bar{s} = \bar{s}, \;\bar{s} \otimes s\bar{s} = s$
\\
Quantum dimension: 1 (for each simple object)
\\
Twists: $\theta(s) = i, \theta(\bar{s}) = -i, \theta(s\bar{s}) = 1$
\\
\\
$\bullet \;\;$ Ising MTC:
\\
Simple $\mathcal{O}bj: \{1 , \sigma, \psi \} $
\\
Fusion: $\sigma \otimes \sigma = 1 \oplus \psi, \; \sigma \otimes \psi = \sigma = \psi \otimes \sigma, \; \psi \otimes \psi = 1 $
\\
Quantum dimension: $\{1: 1,\; \sigma : \sqrt{2},\; \psi: 1\}$
\\
Twists: $ \theta(1): 1, \;\; \theta(\sigma) = e^{\frac{ \pi i}{8}}, \;\; \theta(\psi) = -1  $
\subsubsection{$\mathbb{Z}_N$ Toric Codes}
Toric Code ($\mathbb{Z}_2$) can be generalized to $\mathbb{Z}_N$. The Hamiltonian is given by:
\begin{equation} \label{Z_4 TC Hamiltonian}
    \mathcal{H} = -\sum_{v}A_v - \sum_p B_p - H.c.
\end{equation}
where H.c. corresponds to hermitian conjugate. Let $w$ be an n-th root of unity. Then, $A_v$ and $B_p$ are defined by:
\begin{figure}[h]
    \centering
    \includegraphics[width = 0.5\textwidth]{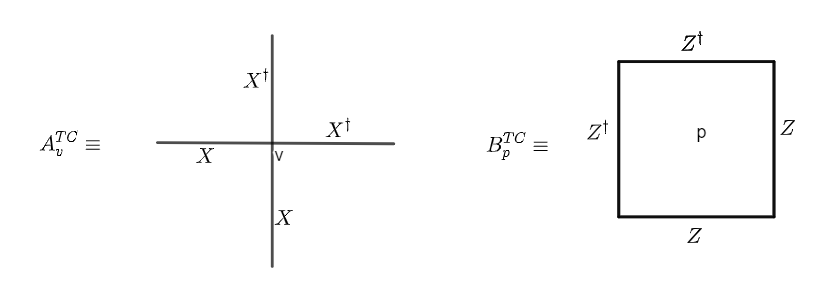}
    \caption{$\mathbb Z_N$ Toric code operators}
\end{figure}
\begin{equation} \label{Z_N TC pauli operator}
    X = \sum_{k \in \mathbb Z_N} |k+1\rangle\langle k| \;\;\;\;\;\; Z = \sum_{k \in \mathbb Z_N} \omega^k |k\rangle\langle k|
\end{equation}
The dimension of the logical subspace is $N^2$. The calculation is similar to the regular TC. For H horizontal shifts and V vertical shifts, there are HV vertex and HV plaquette operators of order N and 2HV number of N-qudits. There are two trivial constraints $\prod_v A_v = 1 = \prod_p B_p$. Therefore, the logical space dimension is $N^{2HV}/N^{HV + HV -2} = N^2$
\\
The excitations are simple objects mentioned in the previous section, a mixture of electric charges and magnetic fluxes. $\mathbb{Z}_N$ TC is a little more exciting as it gives more freedom for excitations and room to condense certain anyons as described later.

\section{Topological Twists}
\subsection{Bombin's Lattice}
Here we recall Bombin's twists on Toric code lattice \cite{Bombin2010} and some modifications as well as possible generalizations. One can define a Toric code where there is one type of operator and the qubits are on the vertices. The Hamiltonian is defined as:
\begin{equation} \label{Bombin's Hamiltonian}
    \mathcal{H} = -\sum_k A_k \;\;\;\;\;\;\;\;\;\;\;\;\; A_k := \sigma_k^x \sigma_{k+i}^z \sigma_{k+i+j}^x \sigma_{k+j}^z
\end{equation}
where k runs through all plaquettes. Here k = (i, j) indexes the spins and i:= (1, 0), j:=(0, 1). Let $|A_k|$ be the eigenvalue of the operator $A_k$.  The ground state is given by $|A_k| = 1$ for all k. Excitations correspond to $|A_k| = -1$ for some k and are gapped.
\begin{figure}[h]
    \centering
    \includegraphics[width = 0.35\textwidth]{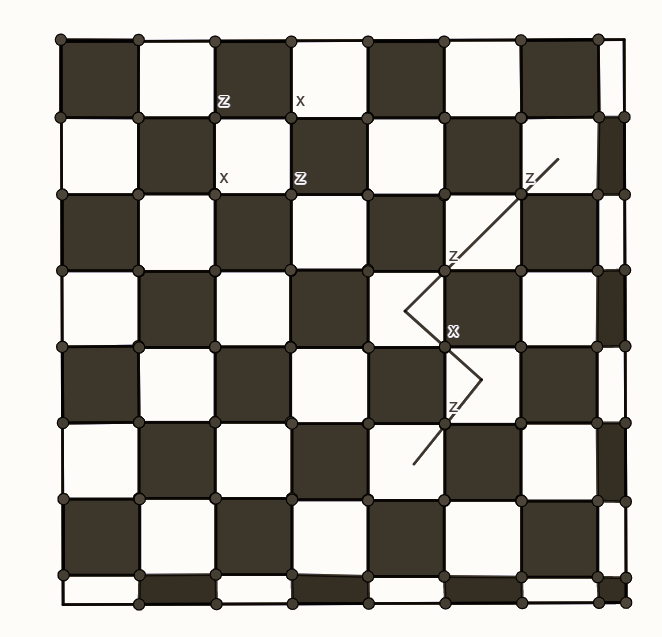}
    \caption{Toric code with qubits on the vertices}
\end{figure}
\\
\textbf{Dimension counting}:
Suppose there are n vertices on each horizontal line and m vertices on each vertical line, then the total number of qubits is nm. There are nm number of plaquettes (operators) as well. There are two trivial operators. All the dark plaquettes' product is 1 and so is the light plaquettes' products. Therefore the logical dimension is $2^{nm}/ 2^{nm-2} = 2^2 = 4$ which agrees with the results of Toric code.
\\
\textbf{String operators}:
There are two types of excitations, corresponding to the dark (e, charge) and light (m, flux) plaquettes. These quasi-particles (excitations) correspond to the quantum double $D(\mathbb Z_2)$. Since the light and dark coloring is a choice, there is a trivial symmetry between the e and m labels.
A $\sigma^z$ on the bottom left corner of a plaquette creates two plaquettes (diagonal to each other) operators anti-commute, therefore a pair of quasiparticles are created. The string operator $S_{\gamma}$ is a concatenation of ordered pair of segments $S_{\gamma} = S_{t_k} \dots S_{t_1}$ where $t_i$ are different plaquettes and each $S_{t_i}$ flips the eigenvalues of the plaquettes at the end of the string.
\\
These operators have two properties:
\\
i) The light $S_{\gamma^L}$ and dark $S_{\gamma^D}$ string operators anti-commute.
\\
ii) String operators can be freely deformed as long as they don't go over any excitation.
\\
If a closed $S_{\gamma^D} = 1$, then the total charge is $\bm{1}$ or $e$. If a closed $S_{\gamma^L} = 1$, then the total charge is $\bm{1}$ or $m$. Assigning value to $S_{\gamma}$ means it's eigenvalue. Total charge denotes the total excitation in the enclosed region (up to parity).

\subsection{Twist}
A twist is a line defect that changes the lattice structure along a line, therefore changing the geometry which leads to a change in the Hamiltonian.
Five plaquettes are removed and four new ones are introduced, out of which two are pentagons and two parallelograms. The pentagons have a Y at the qubit which is a trivalent vertex, and the rest is similar to a regular square, $\sigma^z$ on the diagonal and $\sigma^x $ on the anti-diagonal. Now a string operator passing through the twist does not close, i.e., a light string operator changes to a dark string operator once it passes through the twist region. This means, that $e$ charges become $m$ fluxes and vice versa when moving through the defect. These twists have a topological nature, they can not be destroyed locally. Therefore, one needs to wind the strings twice through the twists to get a closed string.
\\
Locally, in the twist region, $em$ is condensed, which means, one can get $em$ out of the vacuum in the twist region (more like a line). The twist acts like a functor that sends $e$ to $m$ and vice versa.
\\
\\
The existence of both $e$ and $m$ particles together makes the system non-abelian locally. Ising anyons can be synthetically created \footnote[2]{The pair of e and m due to the twist behaves like Ising anyons}. It's possible to characterize braid relations with the help of Majorana operators. These are self-adjoint, and satisfy $c_j c_k + c_k c_j = 2\delta_{jk}$. The total charge of the j-th and j+1-th anyon is given by $-ic_j c_{j+1}$, the eigenvalues +1 and -1 corresponding to the total charge 1 and $\psi$ respectively. These operators behave the following way through braiding:
    \begin{equation}\label{majorana operators}
        c_j \rightarrow c_{j+1}, \;\;\;\;\;\;\;\; c_{j+1} \rightarrow -c_j, \;\;\;\;\;\;\;\; c_k \rightarrow c_k
    \end{equation}
where $j\neq k \neq j+1$.
\\
Due to the twist, string operators need to go through the twist twice to close off. One can also annihilate the $e$ and $m$ charge.
\begin{figure}[h]
    \centering
        \includegraphics[width = 0.4\textwidth]{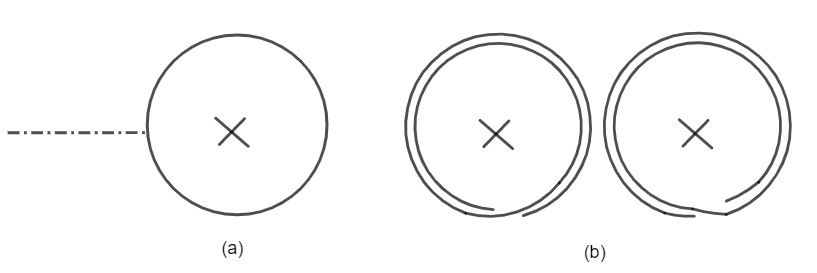}
        \caption{A free fermion in the system and Ising anyons}
\end{figure}
\\
The dotted lines in part (a) represent a pair of e and m.
Part (b) describes two types of Ising anyons $\sigma^+, \sigma^-$ created depending on the braiding between e and m.
Twists act as sources and sinks for $\epsilon$ fermions. The string $\gamma$ winds twice around the twist. The eigenvalues of string operator $S_{\gamma}$ = $\pm i$, corresponding to $\sigma^{\pm}$ particles.
There are six charges or excitations in the system now. $\bm{1}, e, m, \epsilon, \sigma^+, \sigma^-$ which follow fusion rules:
    \begin{equation} \label{ising_fusion}
        \sigma^{\pm} \otimes \sigma^{\pm} = \bm{1} \oplus \epsilon, \;\;\;\; \sigma^{\pm} \otimes \sigma^{\mp} = e \oplus m, \;\;\;\; \sigma^{\pm} \otimes \epsilon = \sigma^{\pm}, \;\;\;\; \sigma^{\pm} \otimes e = \sigma^{\pm} \otimes m = \sigma^{\mp}
    \end{equation}
Associate $\sigma^+$ to $\sigma$ and $\epsilon $ to $\psi$ and the braiding is given by string operators that realize the Majorana operators. However, there is a subtle detail to notice. They behave like Ising but not actually are. $\sigma^+$ can be associated to $e + m$, which makes $\sigma^- = 1 + \epsilon$ from the equations above. But since $(\sigma^+)^2 = 1 + \epsilon$, $\sigma^-$ can not be $(\sigma^+)^2$. The Ising category has twist statistics as:
\begin{equation}
    \theta(\sigma^+) = e^{\pi i/ 8}, \theta(\psi) = -1
\end{equation}
However, one can not get 16th root of unity as the twist statistics from $\mathbb{Z}_2$ model. Therefore, this model realizes ``Ising type" behavior and not the doubled Ising behavior.
\\
Another reason why one can not create a doubled Ising from Toric code is that the twist operation is local, therefore the total quantum dimension of the system remains the same. $\mathbb{Z}_2$ TC has Q. dim 2, and Ising has Q. dim 2, whereas doubled Ising has Q. dim 4.

\subsection{Translating Bombin's lattice to Kitaev's lattice}
The lattice with qubits on the vertices can be translated to Kitaev's lattice where the qubits are on the edges, to get a better picture of what happens at the twist, with the ability to generalize it further.
\begin{figure}[h]
    \centering
    \includegraphics[width = 0.3\textwidth]{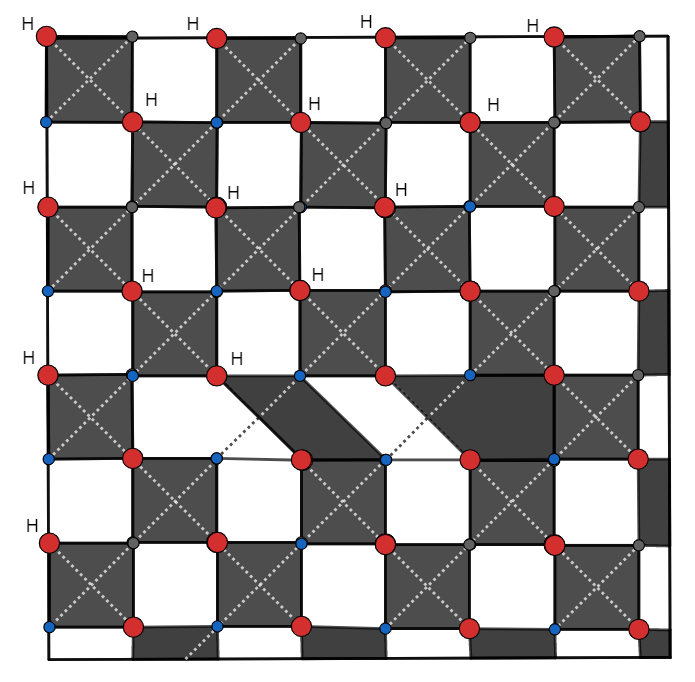}
    \includegraphics[width = 0.3\textwidth]{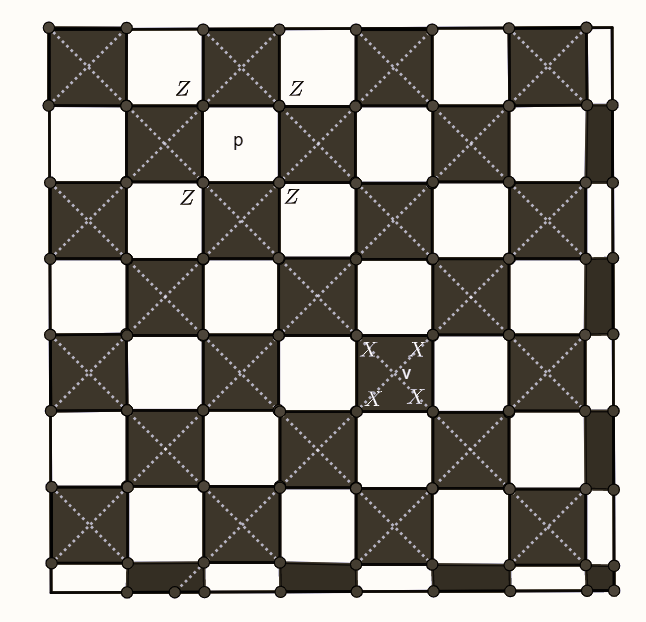}
    \includegraphics[width = 0.285 \textwidth]{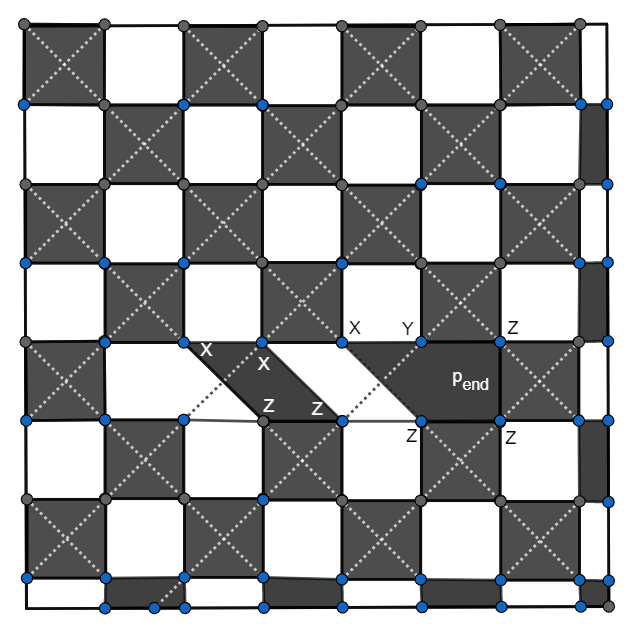}
    \caption{(i) Applying Hadamards to change Bombin's lattice to Kitaev's lattice, (ii) Realizing the plaquette and the star operators, (iii) Realizing the twist operators in Kitaev's lattice}
\end{figure}
\\
Apply Hadamard operators on the diagonal vertices which exchange the Z operator to X and vice-versa. For the new lattice, the intersection of the dotted white lines is the vertices which sit in the dark squares. After applying Hadamard, the diagonal operators change from Z to X, so each operator is X for the vertex. Similarly, for the plaquette, the anti-diagonals operators change from X to Z.
\begin{figure}[h]
    \centering
    \includegraphics[width = 0.315\textwidth]{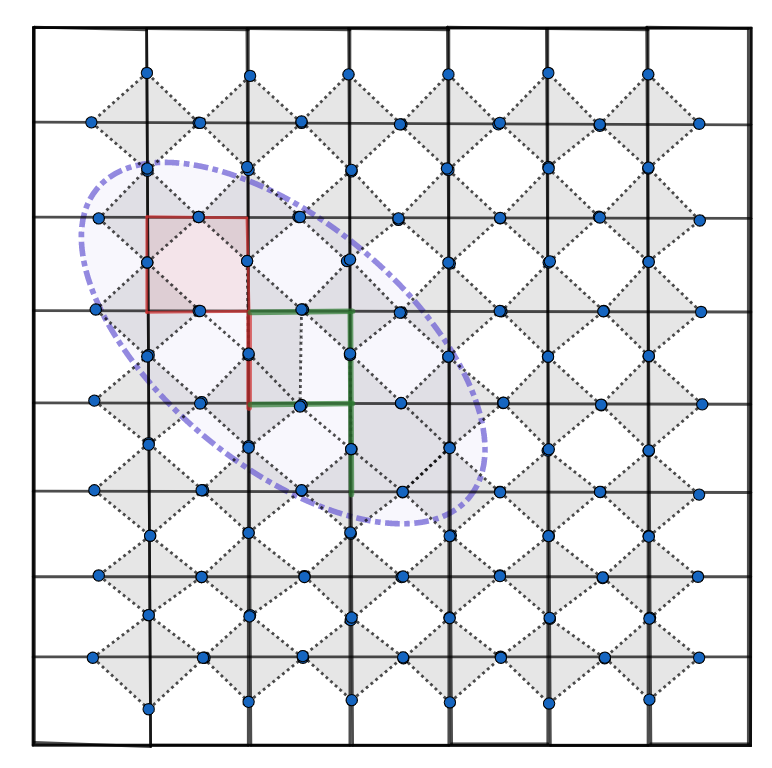}
    \includegraphics[width = 0.35\textwidth]{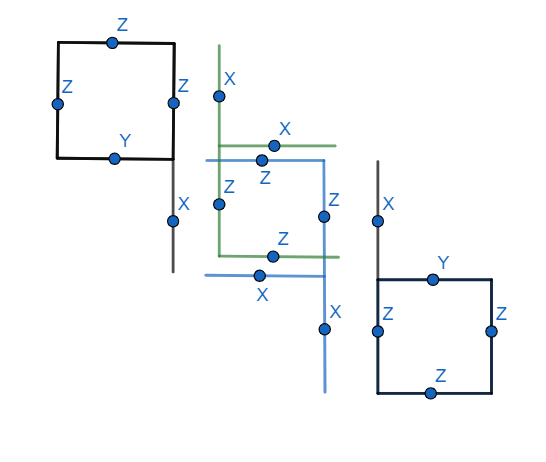}
    \includegraphics[width = 0.315\textwidth]{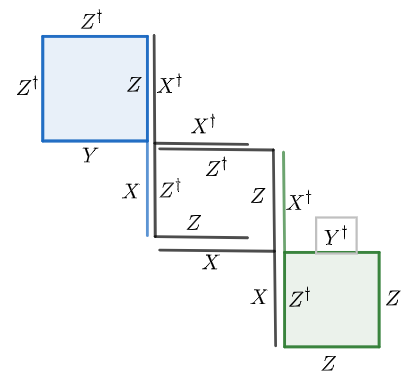}
    \caption{(i) Embedding Bombin's lattice into Kitaev's, (ii)Twist operators for $\mathbb Z_2$ and (iii) $\mathbb Z_N$ }
\end{figure}
\\
The twisted region is highlighted.
The vertex and plaquette operators are broken into halves and glued differently to give the required twist to the lattice. The inverted/reflected F operators correspond to the switching of type e particle to type m particle or the other way around. The twist region can generate a pair of e and m or can absorb the pair. Note that it's possible to extend this notion of twist to any $\mathbb{Z}_N$ TC model because of this representation.
\\
Similar to the previous calculations, the dimension after this twist in $\mathbb{Z}_N$ Toric Code is the same as $\mathbb{Z}_4$ TC, i.e. $N^2$.
\subsection{Twisted figures}
These twists have some significance, but what makes them topological? The confined (contractible) twists correspond to additional (local)patches on the torus, which changes the topology slightly, but not quite.
\begin{figure}[h]
    \centering
    \includegraphics[width = 0.75\textwidth]{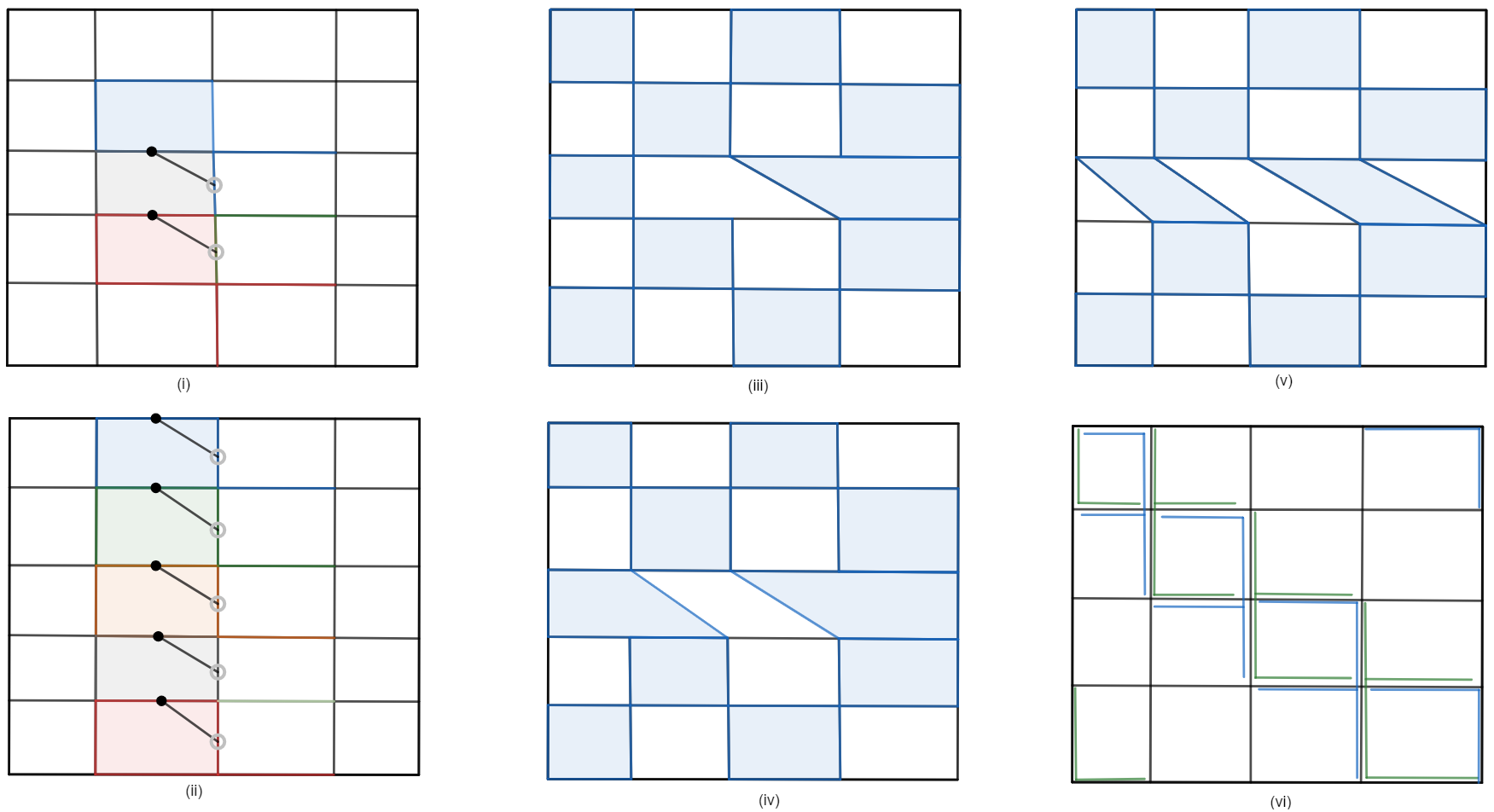}
    \caption{Different types of twists}
    \label{fig: Twists}
\end{figure}
\\
Fig \ref{fig: Twists} (i) and (ii) correspond to twists mentioned in \cite{Krishna2020}. (i) preserves the logical dimension of the manifold. Three plaquettes and three star operators change into three fish and two short operators. The single line with a black dot and grey annulus represents an XZ operator, with X being the black dot and Z being the grey annulus. The number of constraints is reduced by 1, however, there is only a trivial constraint instead of 2 now. In (ii), because of the periodicity, the number of operators remains the same, however, the number of trivial constraints is decreased from two to one. Therefore, the logical dimension decreases from 4 to 2.
\\
(iii), (iv), and (v) correspond to twists in \cite{Bombin2010}. (iii) has the same logical dimension (discussed earlier). The logical dimension remains the same in (iv) as well, because the number of operators is reduced by one as well as the number of trivial constraints. Thus no change in the logical dimension. In (v), the number of operators is the same as before, and the number of trivial constraints change (like (ii)). The logical dimension reduces to two. Fig (ii) and (v) are almost identical.
\\
Fig (vi) corresponds to Bombin's twist in Kitaev's lattice. Notice here that the twist corresponds to an actual twist on the Torus. In this picture, it's clear that there is no horizontal or vertical closed loop (unless it goes around twice).
In Fig (ii), (v), and (vi), the dimension two logical space corresponds to one copy of su(2). However, in (i), (iii), or (iv), the logical dimension is four, but it's slightly more interesting now. The previous copies of su(2) still seem to exist alongside some other logical operators passing through the twist. However, if looked carefully, a product of two such operators can be obtained from the product of two fermions or the four logical operators. If the number of twists is even (given they are separated or have some gap between them), then the even number of Majorana or Ising anyons makes up the logical operator independently.
\\
\begin{figure}[h]
    \centering
    \includegraphics[width = 0.5\textwidth]{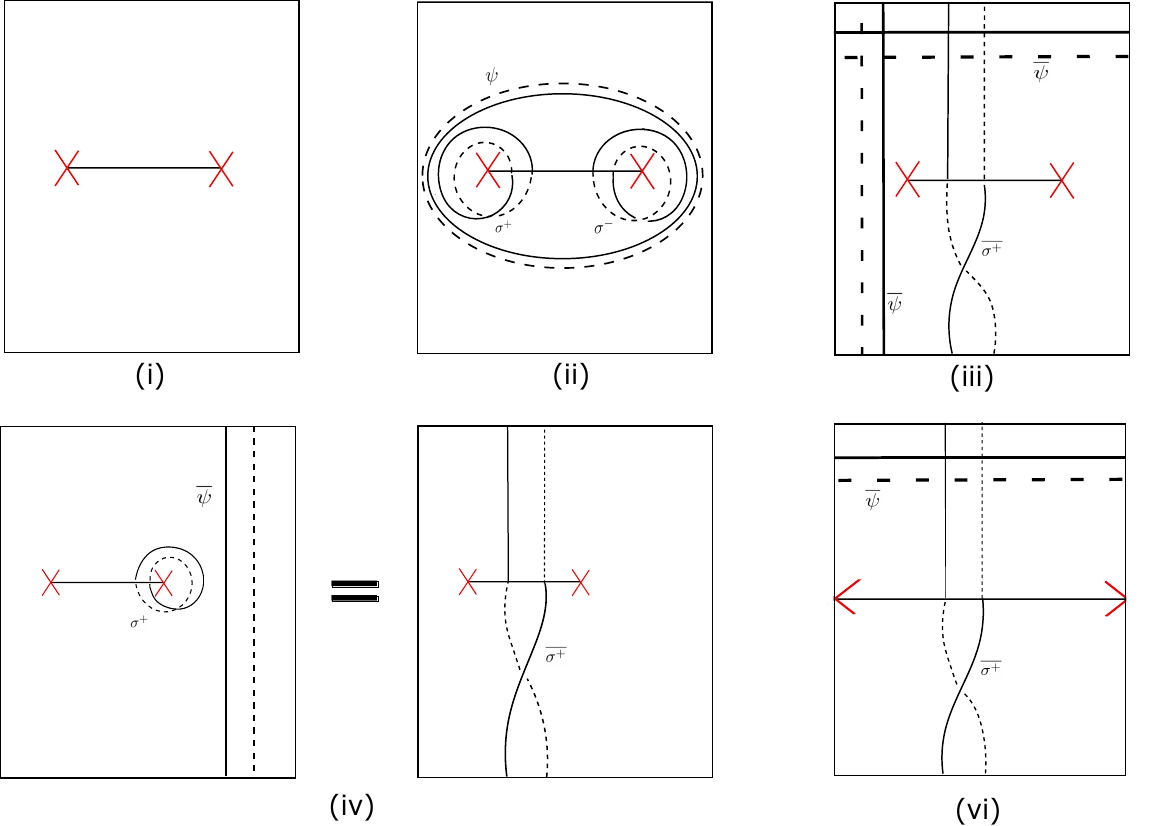}
    \caption{Single Ising twist logical properties }
    \label{fig:ising_prop}
\end{figure}
\\
In fig \ref{fig:ising_prop}, we look at different non-contractible loops corresponding to different particle types. Each non-contractible loop has a bar on top. (i) is just the Ising twist described earlier. (ii) $\sigma^+, \sigma^-, \psi$ particle types. The $\psi$ loop that goes around the twist can be generated through two logical(vertical) $\psi$ outside the twist, which can cancel out with each other, therefore is not logical. (iii) represents non contractible $\sigma$ and $\psi$. $\psi$ can also be considered as the logical Y operator from the Toric code. In (iv), a $\sigma^+$ particle (tensored) with a non-contractible $\psi$ makes up the non-contractible $\sigma$. In the case of a non-contractible twist, there is only one type of non-contractible $\psi$ and a non-contractible $\sigma$. Note that this $\psi$ can now be annihilated when passed through the twist, which means it's trivial. Therefore, the logical operator in this case would be $\bar{Z}$ which is the same class as $\bar{X}$ in this particular case.

If there are two full twists next to each other, it results in logical dimension four (since twists cancel each other out). In fact, in that case, the logical space corresponds to su(2) $\times$ su(2), like the TC. In an odd number of non-contractible twists, the logical dimension decreases to two.

\section{Error Correction/Detection algorithms}
Error correction in surface and toric codes has been studied very well \cite{Bravyi98}. There have been some results studying the string operators in the doubled semion case \cite{Dauphinais19, Levin2005}. However, it either involves a lot of computation or it's difficult to verify the properties of the doubled semion from the string operator easily. Recently, \cite{Ellison2022} studied the doubled semion model as a part of the general Pauli Stabilizer model and generalized results for more condensed toric codes. "All Topological Quantum Doubles with Abelian anyons can be obtained from decoupled copies of $\mathbb{Z}_N$ Toric Code through condensation". They do not mention any algorithm to do error detection or correction explicitly. In this section, we propose the error correction method and its possible applications in TQC.
\\
\textbf{Pauli Stabilizer model}
\\
The Hamiltonian is defined on a square lattice with a 4-qudit on each edge. The Pauli-X and Pauli-Z operators now are of the form
\begin{equation} \label{Z_4 pauli operators}
    X_e = \sum_{j \in \mathbb{Z}_4} |j+1 \rangle \langle j|, \;\;\;\;\; Z_e = \sum_{j \in \mathbb{Z}_4}i^j |j\rangle \langle j|
\end{equation}
In terms of matrices, this is how the operators look like:
\begin{equation*}
    X_e = \begin{bmatrix}
    0 & 0 & 0 & 1\\
    1 & 0 & 0 & 0\\
    0 & 1 & 0 & 0\\
    0 & 0 & 1 & 0
    \end{bmatrix}
    , \;\;\;\; Z_e = \begin{bmatrix}
    1 & 0 & 0 & 0\\
    0 & i & 0 & 0\\
    0 & 0 & -1 & 0\\
    0 & 0 & 0 & -i
    \end{bmatrix}
\end{equation*}
\begin{equation}\label{Z_4 Pauli relations}
    X_e^4 = 1 = Z_{e'}^4, \;\;\;\;\; X^2 Z^2 = Z^2 X^2, \;\;\;\;\;\; Z_e X_e = i X_e Z_e
\end{equation}
The Hamiltonian for this model has the following form:
\begin{equation}\label{Z_4 TC hamiltonian}
    H_{DS} = - \sum_v A_v - \sum_p B_p - \sum_e C_e + H.c.
\end{equation}
H.c. means hermitian conjugate so the Hamiltonian is hermitian.
\begin{figure}[h]
    \centering
    \includegraphics[width = 0.35\textwidth]{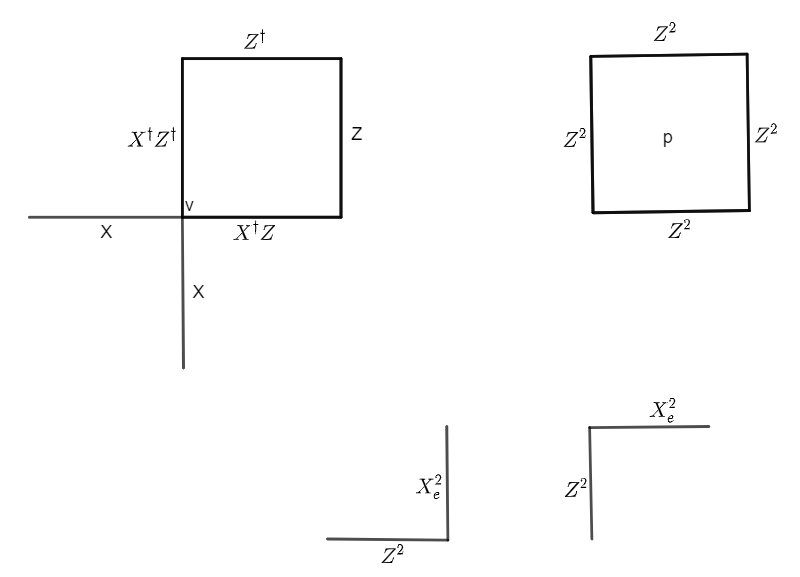}
    \caption{Doubled semion operators}
\end{figure}
\\
All these operators mutually commute. The stabilizer group is
\begin{equation} \label{DS stabilizer}
    \mathcal{S}_{DS} = \langle \{ A_v\},\{B_p\},\{C_e\}\rangle
\end{equation}
\textbf{Excitations}
\\
There are three types of string operators. Two of them are defined on an oriented path on the dual lattice, while one is independent of the orientation and is defined on the lattice.
\begin{equation} \label{DS string operators}
    W_{\bar{\gamma}}^s = \prod_{e \in \bar{\gamma}} W_e^s \;\;\;\;\;\;\;\;\;\;\;\; W_{\bar{\gamma}}^{\bar{s}} = \prod_{e \in \bar{\gamma}} W_e^{\bar{s}} \;\;\;\;\;\;\;\;\;\;\;\; W_{\gamma}^{s\bar{s}} = \prod_{e \in \gamma} W_e^{s\bar{s}}
\end{equation}
The first two create both vertex and plaquette excitations, and the plaquette excitations correspond to $s$ or $\bar{s}$. The third string operator creates $s\bar{s}$ excitation at the vertex (ends of the string).\\
For a given plaquette, there are three string operators $W_{\gamma_1}^a, W_{\gamma_2}^a, W_{\gamma_3}^a$ where $\gamma_1$ is the path that reaches the plaquette from the left, $\gamma_2$ reaches the plaquette from the bottom and $\gamma_3$ reaches the plaquette from the right. Then
\begin{equation} \label{twist from string operators}
    W_{\gamma_1}^a (W_{\gamma_2}^a)^{\dagger} W_{\gamma_3}^a = \theta(a) W_{\gamma_3}^a (W_{\gamma_2}^a)^{\dagger} W_{\gamma_1}^a
\end{equation}
The computation comes out to be:
\begin{equation}\label{DS twists}
    \theta(1) = 1, \;\;\;\;\;\;\;\; \theta(s) = i = -\theta(\bar{s}) \;\;\;\;\;\;\;\; \theta (s\bar{s}) = 1
\end{equation}
This implies $\mathcal{H}_{DS}$ belongs to the DS phase.
If $\alpha$ and $\beta$ are the meridian and longitude of a Torus, then $W_{\alpha}^s = \bar{X_1},\;\; W_{\alpha}^{\bar{s}} = \bar{X_2},\;\; W_{\beta}^s = \bar{Z_1},\;\; W_{\beta}^{\bar{s}} = \bar{Z_2} $ form Pauli X and Pauli Z on the logical subspace $\mathcal{H}_L$. In the logical subspace, these long string operators satisfy:
\begin{align}\label{DS logical operators}
    \bar{X_1}^2 \sim 1 \sim \bar{X_2}^2, \;\;\;\;\;\;\;\; \bar{Z_1}^2 \sim 1 \sim \bar{Z_2}^2 \notag
\\
    \bar{Z_1}\bar{X_1} = - \bar{X_1} \bar{Z_1}, \;\;\;\;\;\;\;\;
    \bar{Z_2}\bar{X_2} = - \bar{X_2} \bar{Z_2}
\end{align}
\begin{figure}[h]
    \centering
    \includegraphics[width = 0.55\textwidth]{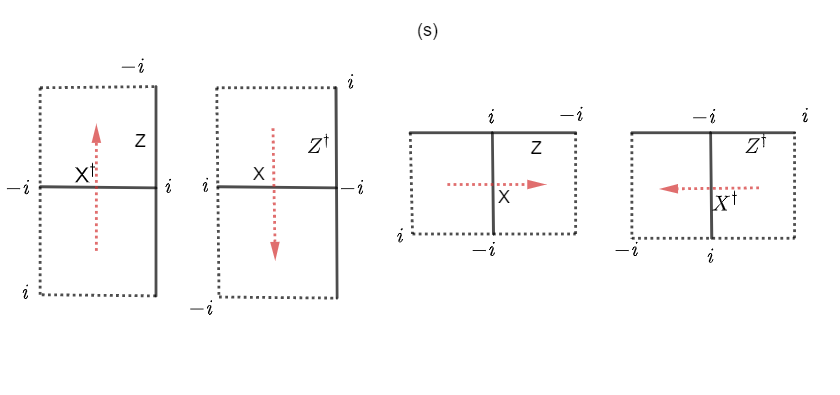}
    \includegraphics[width = 0.55\textwidth]{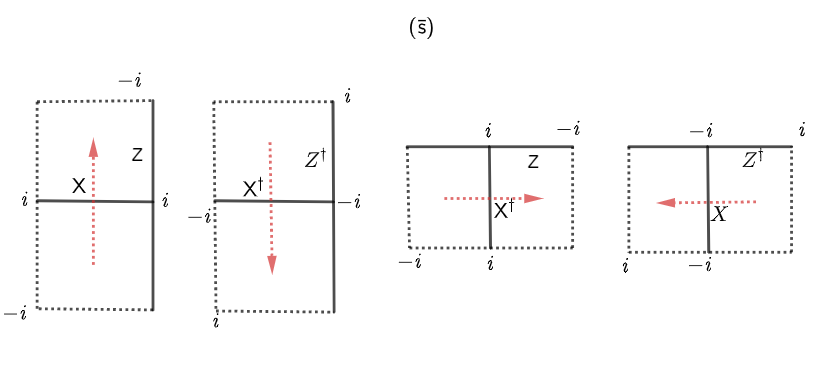}
    \caption{Excitations due to the short string operators}
\end{figure}
\\
There are two excitations on the vertices of one end of the string and two on the other end.
\begin{figure}[h]
    \centering
    \includegraphics[width = 0.215 \textwidth]{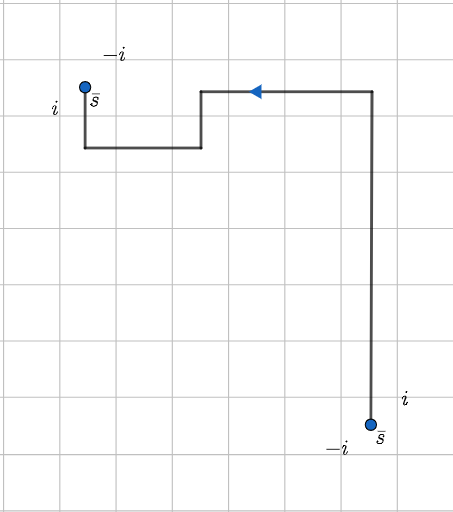} \;\;\;\;\; \includegraphics[width = 0.215\textwidth]{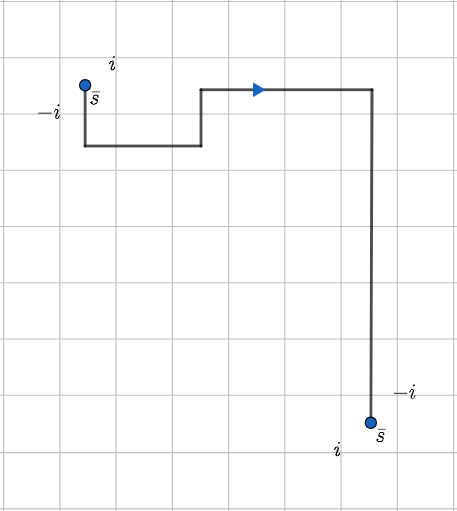}
    \\
    \includegraphics[width = 0.215\textwidth]{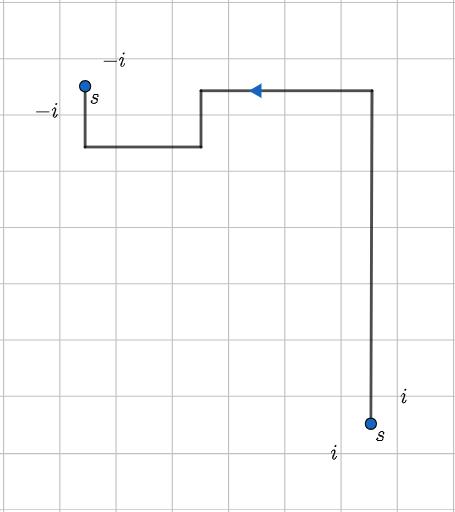} \;\;\;\; \includegraphics[width = 0.215\textwidth]{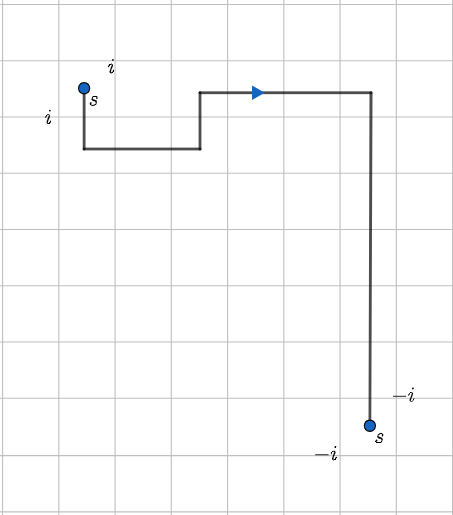}
    \caption{Excitations at the end of string operators}
\end{figure}
\\
There are different ways to close the strings, to get rid of plaquette excitations. If the same string type is closed to form a complete loop, there is no excitation (as expected). But if it's closed with a different string type (half of the string is of type s and the other half of the type $\bar{s}$), then there are two vertex excitations.
If one closes the string in the same direction, then there are four vertex excitations. But if half of the string is of a different type, then one gets $s\bar{s}$ type excitation, which means two vertex excitation.
\begin{figure}[h]
    \centering
    \includegraphics[width = 0.5\textwidth]{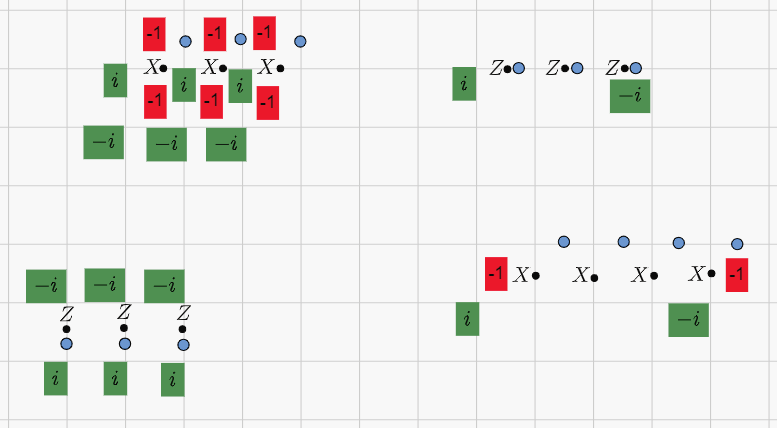}
    \caption{The green values represent vertex excitations, the red ones are the plaquette excitations, the blue dots are the edge excitations and the black dot is the qubit.}
\end{figure}
\\
\textbf{Error correction algorithm:}
\begin{enumerate}
    \item After locating the edge excitations, trace the edge excitations to form a path.
    \item \begin{enumerate}
        \item  If there is a plaquette excitation at one of the ends, apply Pauli X gates to the left of the path.
        \item If there are none, apply Pauli Z operators along the path.
    \end{enumerate}
    \item After locating the plaquette excitation, close it off with with semion string operator (or $\bar{s}$).
    \item  On the two plaquettes where excitations were present, check for vertex excitations on the two corners on the anti-diagonal.
    \item \begin{enumerate}
        \item  If two excitations are present on each plaquette, apply the $s$ string operator twice.
        \item If there is one excitation, apply the $W_{s\bar{s}}$ operator from that vertex to the one on the second plaquette.
    \end{enumerate}
\end{enumerate}

The condensation provides an interesting perspective towards error detection due to the confined particles. Any single bit flip or phase flip error is confined, which means the path they take to propagate increases the energy of the system linearly in terms of the length of the same. This makes it much easier to detect and therefore correct. We will see later an application of the Doubled Semion error detection as being embedded in another system as a patch.

\section{Topological phase transitions}
This section covers the transition between different anyonic theories and an approach to have some of them combined in one single model. The pentagon operators as a part of the Ising patch inside $\mathbb Z_2$ Toric code are the ones that activate the non-abelian properties of the system. The reason is, that a string operator coming in the pentagon operator has an option to go out as a different particle or remain the same even after passing through it. These phase transitions can be characterized by whether something is condensed, which is related to a forgetful functor, or if there is a defect that behaves like the boundary of a functor.
\\
We follow \cite{Kong2014} for notations (mostly).
\subsection{Defects, condensations, and functors}

\subsubsection{Condensation}
\textbf{MTC}: Category $\mathcal{C}$, $\otimes$ the tensor product, $\bm{1}$ the tensor unit object, $\alpha_{X,Y,Z}: (X \otimes Y) \otimes Z \xrightarrow{\simeq} X \otimes (Y \otimes Z) $, braiding $c_{X,Y}: X \otimes Y \xrightarrow{\simeq} Y \otimes X $ and a twist $\theta_X: X \xrightarrow{\simeq} X$. The condensed phases form another MTC $\mathcal{D}$, equipped with tensor product $\otimes_{\mathcal{D}}$, unit tensor $\bm{1}_{\mathcal{D}}$, associator $\alpha_{L,M,N}^{\mathcal{D}}: ( L \otimes_{\mathcal{D}} M) \otimes_{\mathcal{D}} N \xrightarrow{\simeq} L \otimes_{\mathcal{D}} (M \otimes_{\mathcal{D}} N)  $, braiding $C_{L,M}^{\mathcal{D}}: L \otimes_{\mathcal{D}} M \xrightarrow{\simeq} M \otimes_{\mathcal{D}} N $ and a twist $\theta_{M}^{\mathcal{D}}: M \xrightarrow{\simeq} M$
\begin{figure}[h]
    \centering
    \includegraphics[width = 0.35\textwidth]{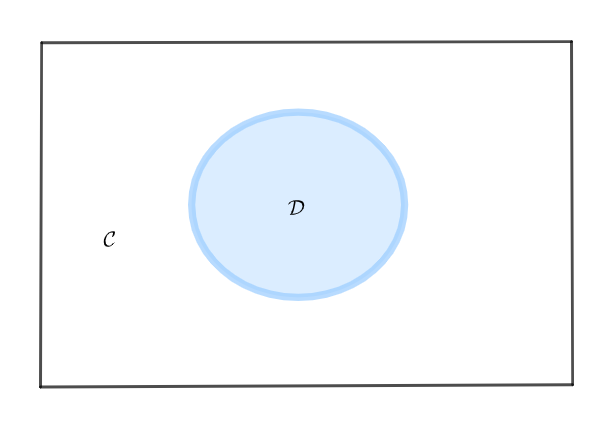}
    \caption{Condensing $\mathcal{C}$ to $\mathcal{D}$}
    \label{fig:categorical condensation}
\end{figure}
\\
Observe that there is a boundary between $\mathcal{C}$ and $\mathcal{D}$ which represents the transfer of bulk properties to wall properties. The boundary is the category $\mathcal{C}_A$, where A is the algebra.
\begin{itemize}
    \item $Obj(\mathcal{D}) \subset Obj(\mathcal{C}) $ with condensation inducing the identity map, $M \xrightarrow{id_M} M$. $\bm{1}_{\mathcal{D}} = A$ where $A \in Obj(\mathcal{C})$. For any non-trivial condensation, $A$ is a composite anyon (direct sum of simple anyons). A is the categorical ground state wave function of the condensed phase.
    \item $hom_{\mathcal{D}}(M,N) \hookrightarrow hom_{\mathcal{C}}(M,N) $, which means $\mathcal{D}$ is (almost) a subcategory of $\mathcal{C}$. There exists a morphism $\iota_A: \bm{1} \rightarrow A$ in $\mathcal{C}$.
    \item There exists an onto \textit{condensation map} $\rho_{M,N}: M \otimes N \rightarrow M \otimes_{\mathcal{D}} N$. There is a canonical morphism $e_{M,N}: M \otimes_{\mathcal{D}} N \rightarrow M \otimes N$ such that $\rho_{M,N} \circ e_{M,N} = id_{M \otimes_{\mathcal{D}} N} $
    \item $A \otimes_{\mathcal{D}} A = A  $ and $ A \otimes_{\mathcal{D}} M = M = M \otimes_{\mathcal{D}} A $. $\mu_{A} = \rho_{A,A}$. $A \otimes A = A \oplus X$ where X is the cokernel of $e_{A}$. This means there is a map $r_X: A \otimes A \rightarrow X$, such that $ r_X \circ e_A = 0$.  Also, $\mu_A \circ e_{A} = id_A$. For unitary categories, $e_{M,N} = \rho_{M,N}^*$
\end{itemize}
\subsubsection{Vacuum in $\mathcal{D}$}
\begin{itemize}
    \item $\mu_A: A \otimes A \rightarrow A $ is an associative multiplication. One gets a commutative \textit{$\mathcal{C}$-algebra} for the triple $(A, \mu_A, \iota_A)$
    \begin{align}\label{algebra morphism}
        \mu_A \circ (\mu_A \otimes id_A) \circ \alpha_{A,A,A} = \mu_A \circ (id_A \otimes \mu_A)
        \\
        \mu_A \circ (\iota_A \otimes id_A) = id_A = \mu_A \circ (id_A \otimes \iota_A)
    \end{align}
    \item The map $A \otimes A \rightarrow A \otimes  A \otimes  A \rightarrow A \otimes  A \rightarrow  X$ is a zero map. $A \otimes X \xrightarrow{1e_X} A \otimes A \otimes A \xrightarrow{\mu_A 1} A \otimes A  \xrightarrow{r_X} X  $ defines a left A-module structure on X. Similarly a right module structure can be defined. $hom_{\mathcal{C}}(\bm{1},A) \cong \mathbb C $. All objects M in $\mathcal{D}$ are A-modules and morphisms are A-module maps. The algebra is separable and connected.
\end{itemize}
\begin{itemize}
    \item A 2d condensable algebra A in MTC $\mathcal{C}$ is a connected commutative symmetric normalized-special Frobenius algebra $\mathcal{C}$.
    \item In TC, anyons form a MTC $Z(Rep(\mathbb Z_2))$, which is the Drinfield center of the fusion category $Rep(\mathbb Z_2)$. $1 \oplus e$ and $1 \oplus m$ are the two 2d-condensable algebra in $Z(Rep(\mathbb Z_2))$. These 2d-condensable algebra can only get one to a trivial subcategory $\mathcal{D}$. This means that the boundary is where certain particles are "absorbed". More precisely, algebra $\bm{1} \oplus m$ corresponds to smooth boundary and algebra $\bm{1} \oplus e$ corresponds to rough boundary \cite{Bravyi98}.
    \item In $\mathbb Z_4$ TC, one example of condensable algebra is
        $1 \oplus e^2 m^2$. The condensation preserves twists, therefore $ id_A \circ \theta_A = \theta_A^{\mathcal{D}} \circ id_A $. But, $\theta_A^{\mathcal{D}} = id_A$, thus, $\theta_A = id_A$.
    Therefore any condensable algebra will be a boson.
\end{itemize}

\subsubsection{Deconfined particles}
Consider the condensation map $\mu_M:= \rho_{A,M}: A \otimes M \rightarrow A \otimes_{\mathcal{D}} M = M$ and $e_{A,M}: A \otimes_{\mathcal{D}} M \hookrightarrow A \otimes M $
\begin{itemize}
    \item $(M,\mu_M)$ is a left A-module, i.e.
    \begin{align}\label{pentagon axiom}
        A \otimes (A \otimes M) \xrightarrow{\alpha_{A ,A,M}} (A \otimes A ) \otimes M \xrightarrow{\mu_A 1} (A \otimes_{\mathcal{D}} A) \otimes M \xrightarrow{\mu_M} (A \otimes_{\mathcal{D}} A) \otimes_{\mathcal{D}}M =M
        \\
        A \otimes (A \otimes M) \xrightarrow{1 \mu_M} A \otimes (A \otimes_{\mathcal{D}} M) \xrightarrow{\mu_M} A \otimes_{\mathcal{D}} ( A \otimes_{\mathcal{D}} M) = M
    \end{align}
    These maps are the same. Also, $ \bm{1} \otimes M \xrightarrow{\iota_A 1} A \otimes M \xrightarrow{ \mu_M} M  $ is the identity map.
    \item $(M,\mu_M)$ is a local A-module: condensation respects the braiding i.e. $A \otimes_{\mathcal{D}} M = M = M \otimes_{\mathcal{D}} A $ and $ c_{M,A}^{\mathcal{D}} = c_{A,M}^{\mathcal{D}} = id_M $. Therefore, $\mu_M \circ c_{M,A} \circ c_{A,M} = \mu_M $
\end{itemize}
The notion of a local A-module is the definition of a "deconfined particle". Right A-modules (or left) that is not local is precisely "confined particle". The boundary between $\mathcal{C}$ and $\mathcal{D}$ in fig:\ref{fig:categorical condensation} describes the confined particles, which corresponds to a right module.
\\
The quantum dimensions in $\mathcal{D}$ can be obtained from those in $\mathcal{C}$ as
\begin{equation*}
    dim_{\mathcal{D}}M = dim_{\mathcal{C}}M / dim_{\mathcal{C}}A
\end{equation*}
\begin{theorem}[\cite{Kong2014}]
If A is a 2d condensable algebra in MTC $\mathcal{C}$, then $\mathcal{C}_A^{loc}$ of local A-modules in $\mathcal{C}$ is also modular.
\end{theorem}

\subsubsection{$\mathbb Z_4$ TC}
Consider the category corresponding to $\mathbb Z_4$ TC with 16 anyons
$$\{\bm{1}, e, e^2, e^3, m,m^2,m^3,em,e^2m,e^3m,em^2,e^2m^2,e^3m^2,em^3,e^2m^3,e^3m^3   \}$$
Consider $A = \bm{1} \oplus e^2m^2 $. Let's see whether this is an algebra. We need morphism $\mu: A \otimes A \rightarrow A$ and $\iota : \bm{1} \rightarrow A$.
\\
Consider $\mu$ as the usual tensor product. $(\bm{1} \oplus e^2m^2) \otimes (\bm{1} \oplus e^2m^2) = 1 \oplus e^2m^2$. And $\iota$ sends $\bm{1}$ to $1 \oplus e^2m^2$. In this case $\alpha_{A,A,A} = id $.
\begin{equation}\label{DS condensation morphism}
    \mu \circ (\mu \circ id_A) \circ \alpha_{A,A,A} = \mu \circ (\mu \otimes id_A) = \mu = \mu \circ (id_A \otimes \mu)
\end{equation}
Similarly, the second condition of being an Algebra is also satisfied. Because of the triviality of the tensor product, it's easy to check that it's a connected commutative algebra. $e^2m^2$ is a boson, therefore it preserves twists. Let's try to look into the deconfined particles.
\\
The boundary fusion category is given by $\mathcal{C}_A$. The bulk to boundary map is given by the functor: $-\otimes A: \mathcal{C} \rightarrow \mathcal{C}_A $. Under this functor, we have:
\begin{align}\label{DS condensation functor}
    1 \rightarrow 1 \oplus e^2m^2, \;\;\;\;\;\;\;\;\;\; e \rightarrow e \oplus e^3m^2,& \;\;\;\;\;\;\;\; e^2 \rightarrow e^2 \oplus m^2, \;\;\;\;\;\;\;\;\; e^3 \rightarrow e^3 \oplus em^2 \notag
    \\
    m \rightarrow m \oplus e^2m^3, \;\;\;\;\;\;\;\;\;\; em \rightarrow em \oplus e^3m^3,& \;\;\;\;\;\;\;\; e^2m \rightarrow e^2m \oplus m^3, \;\;\;\;\;\;\;\;\; e^3m \rightarrow e^3m \oplus em^3\notag
    \\
    m^2 \rightarrow m^2 \oplus e^2, \;\;\;\;\;\;\;\;\;\; em^2 \rightarrow em^2 \oplus e^3,& \;\;\;\;\;\;\;\; e^2m^2 \rightarrow e^2m^2 \oplus 1, \;\;\;\;\;\;\;\;\; e^3m^2 \rightarrow e^3m^2 \oplus e\notag
    \\
    m^3 \rightarrow m^3 \oplus e^2m, \;\;\;\;\;\;\;\;\;\; em^3 \rightarrow em^3 \oplus e^3m,& \;\;\;\;\;\;\;\; e^2m^3 \rightarrow e^2m^3 \oplus m, \;\;\;\;\;\;\;\;\; e^3m^3 \rightarrow e^3m^3 \oplus em
\end{align}
$e^2 m^2$ is mapped to the vacuum of the boundary.
\\
There are 7 simple right $A$ modules other than A itself,
$$e \oplus e^3m^2,\; e^2 \oplus m^2,\; e^3 \oplus em^2,\; e^2 m^3 \oplus m,\; em \oplus e^3m^3,\; m^3 \oplus e^2m,\; e^3m \oplus em^3$$.
\\
Out of these seven, only three of them are local modules, namely, $e^2 \oplus m^2,\; em \oplus e^3m^3,\; e^3m \oplus em^3$.
Note that $\theta(e^p m^q) = i^{pq} \implies \theta(e^2m^2) = 1$ and $c_{a,b} \circ c_{b,a} = \theta(ab) / \theta(a) \theta(b)$ \cite{Ellison2022}.
\\
\\
$\theta(e) = 1 \implies c_{e, e^2 m^2} \circ c_{e^2 m^2, e} = \theta(e^3 m^2) = -1$. Similarly $m, e^3m^2,e^2m^3,e^3, em^2, m^3, e^2m$ do not satisfy the condition $\mu_M \circ c_{M,A} \circ c_{A,M} = \mu_M $.
\\
\\
Thus, the UMTC $\mathcal{D} = \mathcal{C}_A^{loc} = \{ 1 \oplus e^2m^2, em \oplus e^3m^3, e^2 \oplus m^2, e^3m \oplus em^3\}$ which corresponds to the Doubled semion topological order.
Since $em$ is not a boson, the Ising twist can not correspond to a 2d-condensable algebra. The 1d defect line does correspond to a functor which sends
\begin{equation}
    1 \rightarrow 1 \oplus em, \;\;\;\;\;\;\;\; e \rightarrow e \oplus m, \;\;\;\;\;\;\;\;\; m \rightarrow e \oplus m, \;\;\;\;\;\;\;\; em \rightarrow 1 \oplus em
\end{equation}
The physical realization of this functor means that at the 1d defect (line), `e' and `m' can be interchanged at the line and an `em' pair lives freely inside the 1d defect line.

\subsection{Simple phase transitions}
\begin{figure}[h]
    \centering
    \includegraphics[width = 0.5\textwidth]{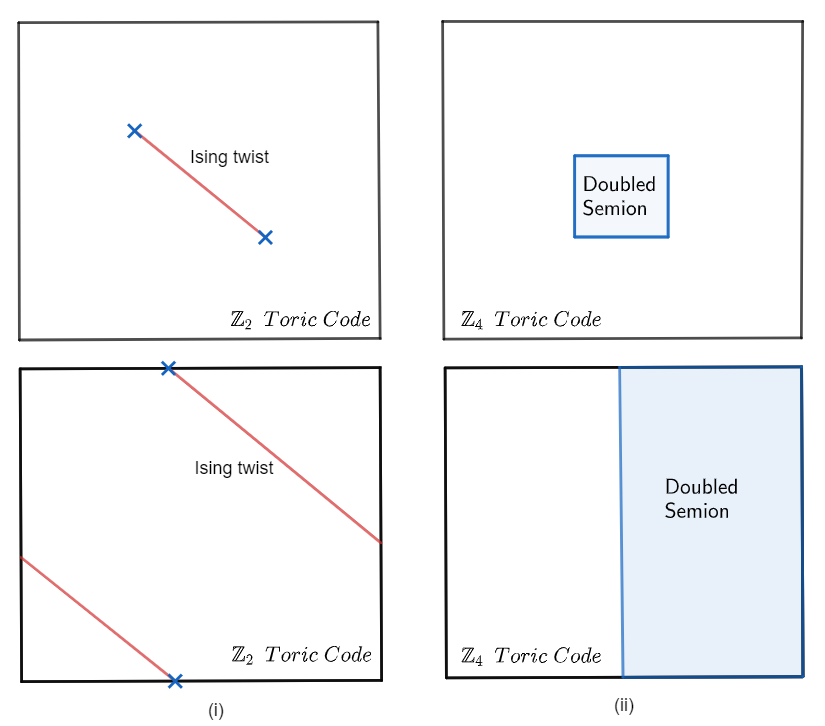}
    \caption{simple phase transitions: (i) Ising twist inside $\mathbb{Z}_2$ Toric code, (ii) Doubled semion model inside $\mathbb Z_4$ Toric code  }
    \label{fig:simple phase transitions}
\end{figure}
In fig \ref{fig:simple phase transitions} part (i), the top is a usual Ising twist. The bottom part is a non contractible twist inside the Toric code. The cross is the pentagon operator. In the non-contractible twist, the pentagonal operator still exists, therefore the model behaves like Ising.
\\
In part (ii), the top figure corresponds to a patch of doubled semion inside $\mathbb{Z}_4 $ Toric code. This patch is slightly different from the way the doubled semion model is created in \cite{Ellison2022}, the reason being, that it doesn't commute with the Toric code operators all around the boundary. More details on the doubled semion patch inside $\mathbb Z_4$ Toric code can be seen in the next figure.
\begin{figure}[h]
    \centering
    \includegraphics[width = 0.4\textwidth]{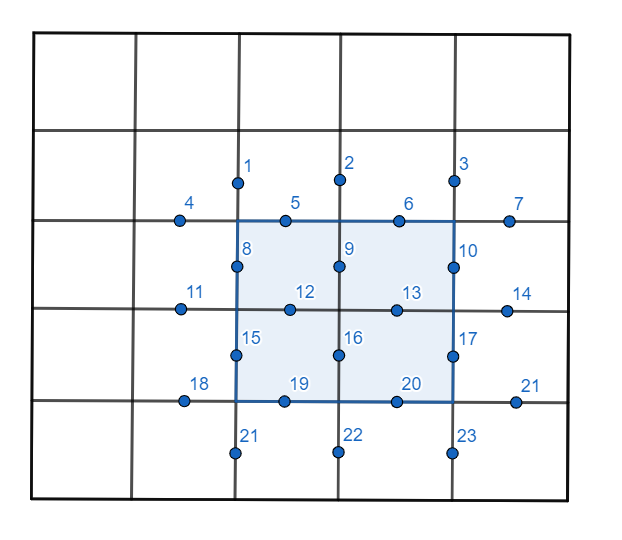}
    \caption{Contractible patch of Doubled Semion inside $\mathbb Z_4$ Toric code}
    \label{fig:Contractible DS patch}
\end{figure}
\\
The operators $A_s^{8,11,12,15}, A_s^{15,18,19,21}, A_s^{16,19,20,22}, A_s^{9,12,16,13}, B_p^{5,8,9,12}, B_p^{6,9,10,13}, B_p^{12,15,16,18}, B_p^{13,16,17,20} $ are removed from the Toric code. Instead, four fish operators, four plaquette operators, and four short string operators are added.
$$ F_{DS}^{8,5,9,12,11,15}, F_{DS}^{15,12,16,19,18,21}, F_{DS}^{9,6,10,13,12,16}, F_{DS}^{16,13,17,20,19,22}, B_{DS}^{5,8,9,12}, B_{DS}^{6,9,10,13}, B_{DS}^{12,15,16,18}, B_{DS}^{13,16,17,20}$$
$$C_{DS}^{9,12}, C_{DS}^{12,15}, C_{DS}^{16,19}, C_{DS}^{13,16} $$
Three fish operators are of order four while $F_{DS}^{9,6,10,13,12,16}$ is of order 2. So, 8 operators of order 4 are replaced by 9 operators of order 2 and 3 operators of order 4. This means the constraint dimension changes by a factor of $4^{-8} 2^{9} 4^3 = 2^{-1}$. Initially, the constraint dimension was $4^N$ where N is the number of 4-qudits. Now it's $2^{-1}4^N$. The trivial constraints now become two-dimensional (two of them).
\begin{equation}
    \prod_p B_p^2 B_{DS} = 1 \;\;\;\;\;\;\;\;\;\;\; \prod A_s^2 F_{DS}^2 B_{DS} = 1
\end{equation}
Therefore the dimension of the logical subspace becomes $4^N / (2^{-1} 4^N) \times 4  = 8 $. This corresponds to an actual phase transition because the dimension of the logical code space decreased from the Logical dimension 16 of $\mathbb Z_4$ Toric code, but is more than the logical dimension of a complete doubled semion model (4). If one makes the patch non-contractible, it's easy to see that the logical dimension of the code space becomes 4, which is the same as the doubled semion. Particularly, the logical operators in this case become the same as the doubled semion logical operators, because they are nothing but a product of logical operators (up to power) of the $\mathbb Z_4$ Toric code.

\subsection{Fun phase transitions}
\begin{figure}[h]
    \centering
    \includegraphics[width = 0.65\textwidth]{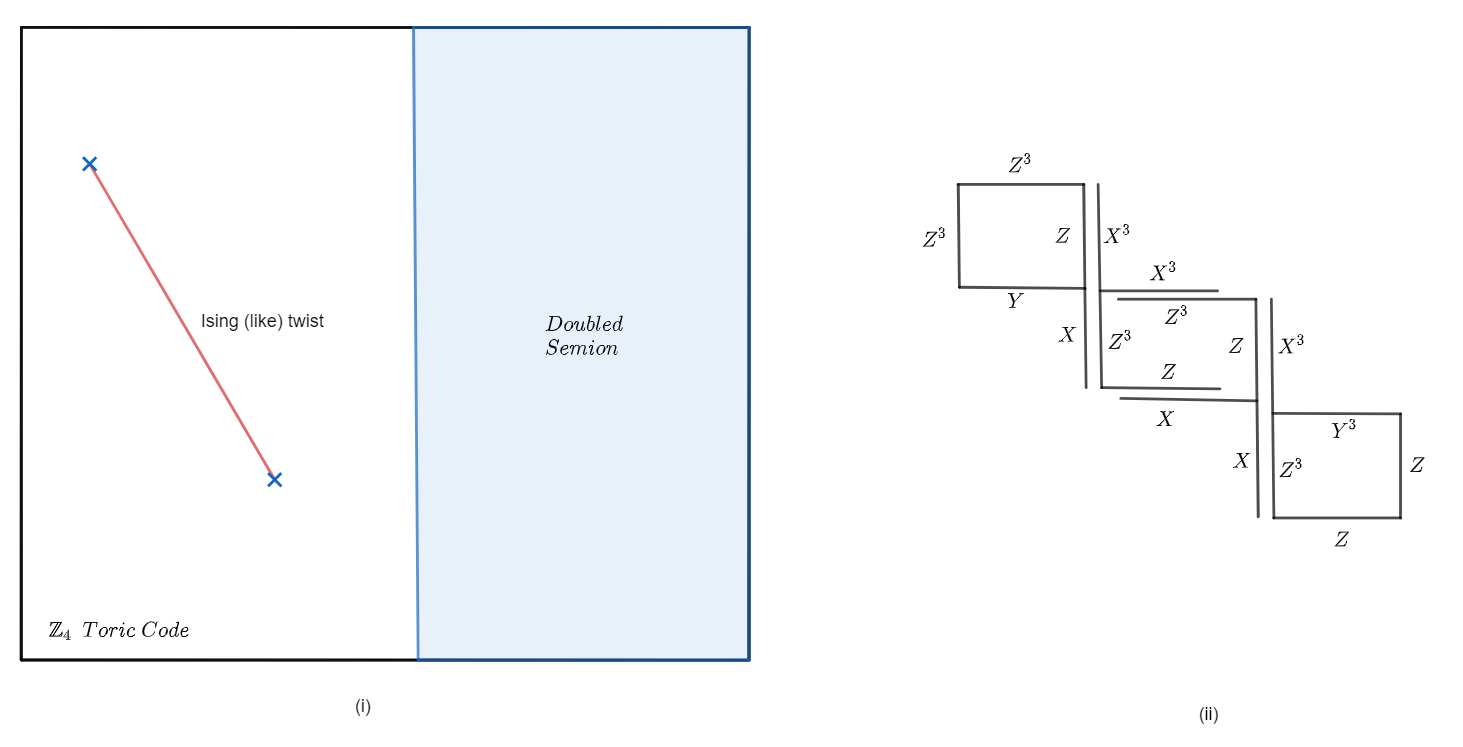}
    \caption{Ising like twist in $\mathbb Z_4$ Toric code and twist operators}
    \label{fig:ising twist in Z_4}
\end{figure}

The phase transitions discussed in the previous subsection relate Toric code to some other model, be it Ising or doubled semion. However, there is another type of phase transition that does not necessarily correspond to any particular (different) theory but is interesting. Ising (like) twists embedded in $\mathbb Z_4$ Toric code fig: \ref{fig:ising twist in Z_4}.
\\
The Ising (like) twist does not change the logical dimension and the logical dimension due to the non-contractible Doubled Semion patch being 4, therefore the logical dimension in the combined picture is also 4. Fig \ref{fig:ising twist in Z_4}(ii) describes the twist in the $\mathbb Z_4$ case, which is very similar to the $\mathbb Z_2$ case. It's possible to have multiple such twists which can lead to interesting effects as the twists exchange e with m or $m^3$ and vice-versa.
\begin{figure}[h]
    \centering
    \includegraphics[width = 0.3\textwidth]{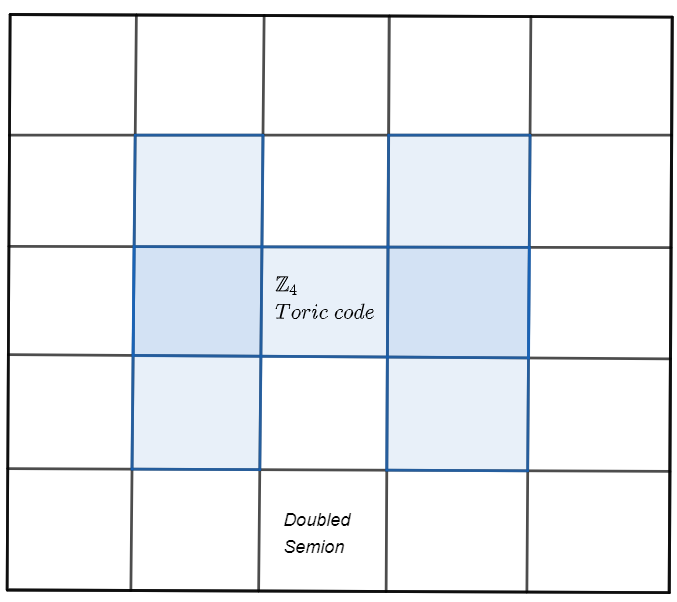} \;\;\;\;
    \includegraphics[width = 0.3\textwidth]{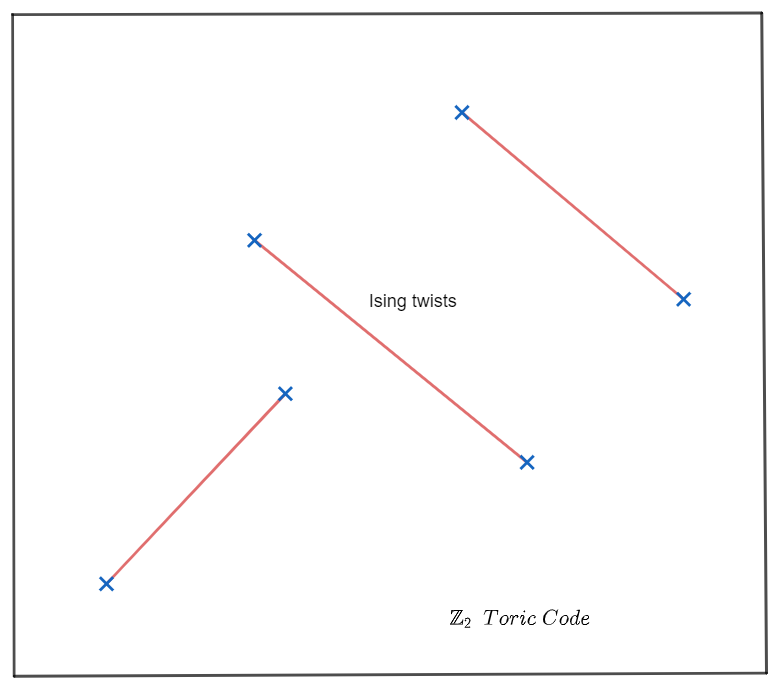}
    \caption{$\mathbb{Z}_4$ in Doubled Semion model and multiple ising twists}
    \label{fig: Z_4 TC in DS}
\end{figure}
\\
Making a $\mathbb{Z}_4$ Toric code patch inside of the doubled semion decreases the logical dimension from 4 to 2. One can have a longer `H' shaped patch which allows particles like $e,\; m, \; e^2m \dots$ to move freely inside the Toric code patch, but they can't escape it without increasing the energy linearly as the traverse.
\\
Similarly, having multiple Ising twists is another interesting defect in Toric code, where the computational power is increased. Each such twist eats up one operator (in terms of number), therefore increasing the logical dimension. By adding a twist, one creates more degrees of freedom and therefore more logical subspace available for computation.
\\
\begin{figure}[h]
    \centering
    \includegraphics[width = 0.7\textwidth]{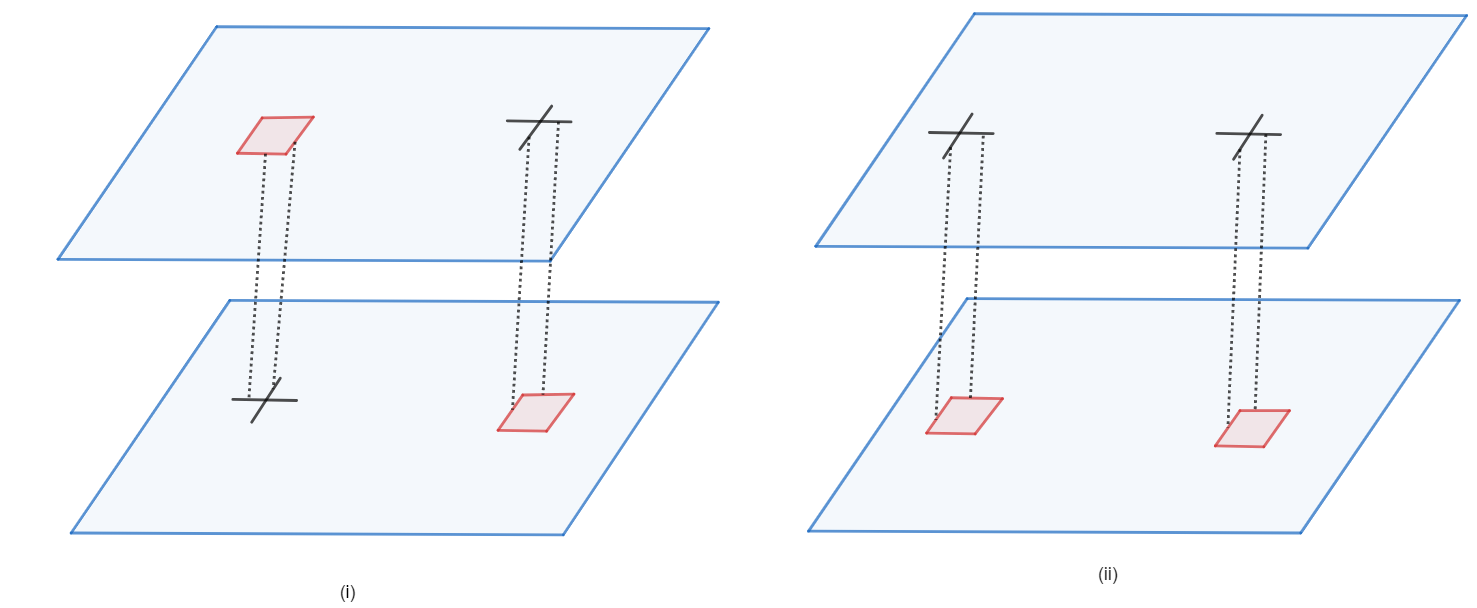}
    \caption{Each parallelogram is a $\mathbb Z_2 $ Toric Code copy. (i) A flux from the upper(lower) TC can become a charge in the lower(upper) TC (ii) A wormhole where fluxes move from the lower TC as charges in the upper TC and vice versa}
    \label{fig:coupled_TC}
\end{figure}
\\
While these twists are on a single lattice structure, it's possible to create a coupled lattice (possibly with twists inside each of them) and create a wormhole. These wormholes have been studied in the single lattice case \cite{Krishna2020}.
\\
In fig \ref{fig:coupled_TC}, (i) two plaquette operators and two-star operators are removed and two bi-layer fish operators are introduced (plaquette in the first TC and star in the second or vice-versa). The dimension of the system remains the same because four constraints are reduced to two, however, the four trivial constraints now merge into two trivial constraints. More precisely, the trivial constraints now are $$\prod B_p^{T1} F_1 \prod A_s^{T2} = 1 = \prod B_p^{T2} F_2 \prod A_s^{T1}$$.
\\
In (ii), a pair of fluxes can go from the lower TC to the upper one and become a pair of charges. Or, a charge in the upper TC and a flux in the lower TC can exist at the same time, which is similar to the Ising twist described earlier. These fish operators are similar to the pentagon operators, which describe an Ising-like behavior. The dimension increases by a factor of 2 because the number of operators (constraints) is reduced by 2, however, there are three trivial constraints now instead of four, $$\prod A_s^{T2} = \prod B_p^{T1} = \prod A_s^{T1} \prod B_p^{T2} F_1 F_2 = 1$$
\\
\\
These ideas of wormholes could help in error detection, error correction, and storing logical information. Different coupling could lead to potentially different non-abelian systems which can help achieve the goal of topological quantum computing faster.
\\
\\
\textbf{Future Plans}: Even though a lot of categories are associated with Topological phases of matter, it's difficult to realize most of them. This paper talks about a way to look into similar phases of matter while keeping the base theory as simple as possible. It would be interesting to see the embedding of different TPMs in $\mathbb Z_N$ Toric codes or coupled toric codes and having non-abelian systems behave as the boundary of a trivially condensed category. While the theory of transition seems easy, there could still be practical hindrances to verifying this behavior and using these models for computation. However, a combination of twisting behavior, different phases of matter in the same lattice, and coupled lattices sound like an approach with a lot of possibilities to explore. There are some similarities between these phase transitions (condensation with a boundary) and condensed matter, namely Andreev's characterization/scattering. I believe this paper could help merge some different areas of physics and mathematics to have more developments in both of them.

\section{References}

\bibliography{main.bib}

\begin{thebibliography}{17}
\providecommand{\natexlab}[1]{#1}
\providecommand{\url}[1]{\texttt{#1}}
\expandafter\ifx\csname urlstyle\endcsname\relax
  \providecommand{\doi}[1]{doi: #1}\else
  \providecommand{\doi}{doi: \begingroup \urlstyle{rm}\Url}\fi

\bibitem[Rowell et~al.(2007)Rowell, Stong, and Wang]{Rowell2007}
Eric~C. Rowell, Richard Stong, and Zhenghan Wang.
\newblock On classification of modular tensor categories.
\newblock \emph{Communications in Mathematical Physics}, 292:\penalty0
  343--389, 2007.

\bibitem[Satzinger(2021)]{Satzinger2021}
et~al. Satzinger, K.~J.
\newblock {Realizing topologically ordered states on a quantum processor}.
\newblock \emph{Science}, 374\penalty0 (6572):\penalty0 1237--1241, dec 2021.
\newblock ISSN 10959203.
\newblock \doi{10.1126/science.abi8378}.
\newblock URL \url{https://www.science.org/doi/10.1126/science.abi8378}.

\bibitem[Acharya(2022)]{Acharya2022}
et.~al. Acharya, Rajeev.
\newblock Suppressing quantum errors by scaling a surface code logical qubit.
\newblock 2022.
\newblock \doi{10.48550/ARXIV.2207.06431}.
\newblock URL \url{https://arxiv.org/abs/2207.06431}.

\bibitem[Bombin(2010)]{Bombin2010}
H.~Bombin.
\newblock {Topological order with a twist: Ising anyons from an Abelian model}.
\newblock \emph{Physical Review Letters}, 105\penalty0 (3):\penalty0 030403,
  jul 2010.
\newblock ISSN 00319007.
\newblock \doi{10.1103/PhysRevLett.105.030403}.
\newblock URL \url{https://link.aps.org/doi/10.1103/PhysRevLett.105.030403}.

\bibitem[Kitaev and Kong(2012)]{Kitaev2012}
Alexei Kitaev and Liang Kong.
\newblock Models for gapped boundaries and domain walls.
\newblock \emph{Communications in Mathematical Physics}, 313\penalty0
  (2):\penalty0 351--373, jun 2012.
\newblock \doi{10.1007/s00220-012-1500-5}.
\newblock URL \url{https://doi.org/10.1007%2Fs00220-012-1500-5}.

\bibitem[Ellison et~al.(2022)Ellison, Chen, Dua, Shirley, Tantivasadakarn, and
  Williamson]{Ellison2022}
Tyler~D. Ellison, Yu~An Chen, Arpit Dua, Wilbur Shirley, Nathanan
  Tantivasadakarn, and Dominic~J. Williamson.
\newblock {Pauli Stabilizer Models of Twisted Quantum Doubles}.
\newblock \emph{PRX Quantum}, 3\penalty0 (1):\penalty0 010353, mar 2022.
\newblock ISSN 26913399.
\newblock \doi{10.1103/PRXQuantum.3.010353}.
\newblock URL \url{https://link.aps.org/doi/10.1103/PRXQuantum.3.010353}.

\bibitem[Kong(2014)]{Kong2014}
Liang Kong.
\newblock {Anyon condensation and tensor categories}.
\newblock \emph{Nuclear Physics B}, 886:\penalty0 436--482, jul 2014.
\newblock ISSN 05503213.
\newblock \doi{10.1016/j.nuclphysb.2014.07.003}.
\newblock URL \url{http://arxiv.org/abs/1307.8244
  http://dx.doi.org/10.1016/j.nuclphysb.2014.07.003;
  10.1016/j.nuclphysb.2021.115607}.

\bibitem[Krishna and Poulin(2020)]{Krishna2020}
Anirudh Krishna and David Poulin.
\newblock {Topological wormholes: Nonlocal defects on the toric code}.
\newblock \emph{Physical Review Research}, 2\penalty0 (2), 2020.
\newblock ISSN 26431564.
\newblock \doi{10.1103/PhysRevResearch.2.023116}.

\bibitem[Delaney()]{Delaney2019}
Colleen Delaney.
\newblock \emph{A categorical perspective on symmetry, topological order, and
  quantum information}.
\newblock PhD thesis, UC Santa Barbara.
\newblock URL \url{https://escholarship.org/uc/item/5z384290}.

\bibitem[Etingof et~al.(2002)Etingof, Nikshych, and Ostrik]{Etingof2002}
Pavel Etingof, Dmitri Nikshych, and Viktor Ostrik.
\newblock On fusion categories.
\newblock \emph{Annals of Mathematics}, 162:\penalty0 581--642, 2002.

\bibitem[Bravyi and Kitaev(1998)]{Bravyi98}
S.~B. Bravyi and A.~Y. Kitaev.
\newblock 1998.
\newblock \doi{10.48550/arXiv.quant-ph/9811052}.

\bibitem[Her(2020)]{Herringer20}
The toric code.
\newblock 2020.

\bibitem[Lin et~al.(2021)Lin, Levin, and Burnell]{Lin2021}
Chien-Hung Lin, Michael Levin, and Fiona~J. Burnell.
\newblock Generalized string-net models: A thorough exposition.
\newblock \emph{Phys. Rev. B}, 103:\penalty0 195155, May 2021.
\newblock \doi{10.1103/PhysRevB.103.195155}.
\newblock URL \url{https://link.aps.org/doi/10.1103/PhysRevB.103.195155}.

\bibitem[Wen(2017)]{Wen2017}
Xiao-Gang Wen.
\newblock Colloquium: Zoo of quantum-topological phases of matter.
\newblock \emph{Rev. Mod. Phys.}, 89:\penalty0 041004, Dec 2017.
\newblock \doi{10.1103/RevModPhys.89.041004}.
\newblock URL \url{https://link.aps.org/doi/10.1103/RevModPhys.89.041004}.

\bibitem[Rowell and Wang(2017)]{Rowell2017}
Eric Rowell and Zhenghan Wang.
\newblock Mathematics of topological quantum computing.
\newblock \emph{Bulletin of the American Mathematical Society}, 55, 05 2017.
\newblock \doi{10.1090/bull/1605}.

\bibitem[Dauphinais et~al.(2019)Dauphinais, Ortiz, Varona, and
  Martin-Delgado]{Dauphinais19}
Guillaume Dauphinais, Laura Ortiz, S~Varona, and M~Martin-Delgado.
\newblock Quantum error correction with the semion code.
\newblock \emph{New Journal of Physics}, 21, 05 2019.
\newblock \doi{10.1088/1367-2630/ab1ed8}.

\bibitem[Levin and Wen(2005)]{Levin2005}
Michael~A. Levin and Xiao-Gang Wen.
\newblock String-net condensation: A physical mechanism for topological phases.
\newblock \emph{Phys. Rev. B}, 71:\penalty0 045110, Jan 2005.
\newblock \doi{10.1103/PhysRevB.71.045110}.
\newblock URL \url{https://link.aps.org/doi/10.1103/PhysRevB.71.045110}.

\end{thebibliography}

\end{document}